\documentclass[twocolumn,pra,showpacs,amsmath,amssymb]{revtex4}
\usepackage{graphicx}
\usepackage{amssymb,amsthm,amsfonts,amstext}

\newcommand{\B}{\mathfrak{B}}

\newcommand{\Hi}{\mathcal{H}}

\newcommand{\Tr}{\mathrm{Tr}}

\newcommand{\ket}[1]{|#1 \rangle}
\newcommand{\bra}[1]{\langle #1 |}

\newcommand{\be}{\begin{equation}}
\newcommand{\ee}{\end{equation}}
\newtheorem{Theorem}{Theorem}

\newtheorem{Definition}{Definition}
\newtheorem{Lemma}{Lemma}
\newtheorem{Corollary}{Corollary}

\usepackage{amsmath}    
\begin{document}

\title{Probing untouchable environment as a resource for quantum computing}
\author{Masaki Owari$^1$, Koji Maruyama$^{1,2}$, Takeji Takui$^2$, Go Kato$^1$}

\affiliation{$^{1}$NTT Communication Science Laboratories, NTT Corporation, Atsugi-Shi, Kanagawa,
243-0198 Japan}

\affiliation{$^{2}$Department of Chemistry and Materials Science, Osaka City University, Osaka,
558-8585 Japan}

\begin{abstract}
When manipulating a quantum system $S$, its surrounding system, or \textit{environment}, $E$
induces unwanted effects. It is mainly due to its vastness and the lack of knowledge about the
Hamiltonian $H_{SE}$ that governs the dynamics inside $E$ and the interaction with $S$. The detail
of $H_{SE}$ is usually extremely hard to identify, since $E$ can hardly be measured or controlled
directly. Nevertheless, here we show that it is possible to probe and control a part of, if not
all, the dynamics involving $E$, within the timescale in which its effective dimension can be seen
finite. That is, we may be able to let a noisy environment work in our favor as a part of quantum
computer.
\end{abstract}
\date{\today}
\maketitle

\section{Introduction}\label{sec:introduction}
The full control of many-body quantum systems is no doubt a key towards the future nano- and
quantum technologies. Among others, the realisation of quantum information processing
\cite{NIELSEN} has been studied intensively as a good test bed of quantum control as well as an
ultimate engineering task that makes full use of quantum mechanical effects
\cite{Warren1993,rabitz2000}. Yet, manipulating quantum states is extremely hard, since
information encoded in quantum states easily leaks out to the environment due to complex and
inevitable interactions with it. In the theory of open quantum systems, an environment is usually
treated as a large bath \cite{OPENQUANTUM,Nitzan2006}, washing away most of its dynamical details,
rather than a quantum object that we can control actively.

Although it is indeed hopeless to have a full control of infinitely large environment, what if we
knew that the system surrounding a small quantum device is finite dimensional. There exist such
compound quantum systems, in which a finite (possibly high) dimensional system $E$ interacts with
a small device $S$ that is directly controllable \textbf{and measurable}. A good example can be
found in the hybrid system of a superconducting qubit (SC) and nitrogen-vacancy (NV) centres in
diamond \cite{Kubo2010,Kubo2011,Zhu2011}. A SC qubit ($S$), which is under control, is coherently
coupled with a finite number ($10^7\sim 10^{12}$) of electron spins ($E$) trapped by NV centres.
Despite a huge number of spins that are waiting to be controlled, we still lack a method, thus any
proposal towards the exploitation of high dimensionality is strongly coveted. The biggest obstacle
to this end would be the acquisition of precise information of the Hamiltonian that governs the
total dynamics. How can we probe the internal, possibly quite complex, dynamics in $E$ by a
limited access through $S$?

In this paper, we demonstrate how this formidable task of identifying the dynamical structure of
the total system can be achieved, provided the dimension of $E$ can be regarded as finite. More
precisely, for a given principal system $S$ and its surrounding system $E$, we will estimate the
parameters of the Hamiltonian $H_{SE}$, which will be sufficient in terms of the indirect control
of $E$ as a \textit{resource} for quantum engineering, such as quantum computation, through $S$.
Throughout the paper, we shall call $E$ the \textit{environment} symbolically, and its dimension
is assumed to be finite, but unknown a priori, as we will formally state later.

Readers may be reminded of the methods of quantum process tomography
(QPT)\cite{Poyatos1997,Chuang1997,NIELSEN,Mohseni2006,DAriano2001,Duer2001,Altepeter2003} as means
to determine all the parameters that characterise a general quantum evolution, namely a completely
positive (CP) map. Nevertheless, QPT is a scheme to estimate the CP map for a quantum system for
which we can prepare a specific state and perform measurements. Thus, the conventional QPT methods
do not reveal the nature of environment, which is beyond the reach of our measurement.

There have also been a series of studies on Hamiltonian identification of a many-body system under
limited access \cite{Burgarth2009,Franco2009,Burgarth2009a,Burgarth2011}. However, all of them
assume that a priori knowledge is available about the system configuration, and the
controllability of the system state, which includes initialisability. In the present analysis, no
particular assumptions as such are made about system structures or the type of interaction, that
is, the task is even more nontrivial than existing tomographic schemes.

\begin{figure}
\includegraphics[scale=0.8]{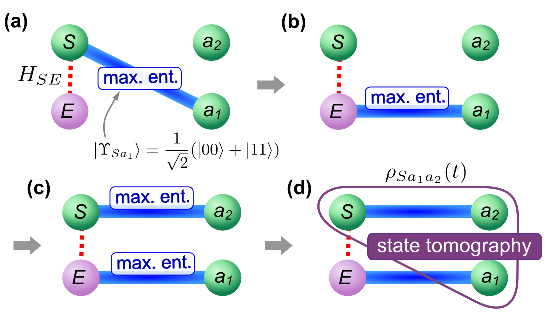}
\caption{A simple example to explain the idea. With the help of ancillary systems, the entire
state can be steered to be two pairs of maximally entangled states as in (c). State tomography of
$\rho_{Sa_1 a_2}(t)$ provides us with information on the Hamiltonian
parameters.}\label{fig:example}
\end{figure}

The identification method we present here consists of two major parts: one is a state-steering
protocol to establish entanglement between $SE$ and an ancillary system $A$, and the other is a
tomographic process to reconstruct $H_{SE}$. Let us depict the basic idea by a simple example with
an illustration in Fig. \ref{fig:example}, whose processes (a)-(c) correspond to the
state-steering protocol and the information of $H_{SE}$ is extracted in (d). Suppose both $S$ and
$E$ are two-dimensional systems. By preparing a maximally entangled state between $a_1$ and $a_2$,
$\ket{\Upsilon_{a_1 a_2}}=(\ket{00}+\ket{11})/\sqrt{2}$, as an ancilla and swapping $S$ and $a_2$,
we can entangle $S$ and $a_1$ as in Fig. \ref{fig:example}(a). We will show later that we
can probabilistically make the evolution of $SE$ be effectively a SWAP
operation, i.e., an operation that transfers entanglement between
$A$ and $S$ to that between $A$ and $E$, even if $H_{SE}$ and the initial
state are unknown. This is possible
as long as $H_{SE}$ is an entangling interaction (Fig. \ref{fig:example}(b)), like the standard
Heisenberg spin interaction. We can then entangle $S$ and $a_2$ to have two maximally entangled
pairs (Fig. \ref{fig:example}(c)) (This is possible because both can be measured and controlled).
Because of a property of maximal entanglement,
\begin{equation}\label{mirroring1}
U_{SE}(\ket{\Upsilon_{Sa_2}}\otimes\ket{\Upsilon_{Ea_1}})=V_{a_1 a_2}
(\ket{\Upsilon_{Sa_2}}\otimes\ket{\Upsilon_{Ea_1}})
\end{equation}
holds when $V_{a_1 a_2}=U_{SE}^T$. Thus, in a sense, the effect of $H_{SE}$ will be reflected in
the dynamics on the side of $a_1 a_2$. In our setting, since $S$ is accessible as well as $a_1$
and $a_2$, the reduced density operator $\rho_{Sa_1 a_2}(t)$ can be obtained through state
tomography on these three subsystems (Fig. \ref{fig:example}(d)). The information about $H_{SE}$
will then be acquired by analyzing $\rho_{Sa_1 a_2}(t)$.

Even when there is no a priori knowledge available on the dimensionality of $E$ or the
type of interaction, we can construct a method for the two major processes, as in the above
somewhat simplistic example. There exists a protocol that steers the entire state so that it will
have the entanglement structure as in Fig. \ref{fig:example}(c), and the Hamiltonian $H_{SE}$ can
be identified through tomography of $SA$.

After giving an outline of the main results in Sec. \ref{sec:main_results}, we will define the
setup and the equivalence class with respect to the observable dynamics in Sec. \ref{sec:problem_setting} more rigorously. In Sec.
\ref{sec:pme}, it will be shown that, if the \textit{Maxmailly Entanglement (ME)} condition, which can be
verified by tomography on $SA$,
is satisfied, simply observing the
natural time evolution of the state on $SA$ is sufficient for identifying the equivalence class.
Also, the fulfillment of the ME condition is shown to guarantee the establishment of a state that
is essentially maximally entangled
between $SE$ and $A$ (Sec. \ref{sec:pme} B). We will then discuss a specific method as to how we
can extract information on $H_{SE}$ from the observed data in Sec. \ref{sec:tomography} and present
the state-steering protocol in Sec. \ref{sec:state_steering} that attains the right entanglement
structure. Section \ref{sec:numerical} shows the results of numerical simulations to confirm our
ideas, taking a network of four spins as an example, before summarising in Sec.
\ref{sec:conclusion}.

\section{Main results}\label{sec:main_results}
Let $\mathcal{H}_S$ and $\mathcal{H}_E$ be the Hilbert spaces of the principal system ($S$) and
its environment ($E$), whose dimensions are $d_S$ and $d_E$, respectively. Then, the assumptions
on which we base our analysis are as follows.
\begin{itemize}
\item[(i)] $d_E$ is finite, although its value may be unknown.

\item[(ii)] Ancillary states, each of which is a maximally entangled pair,
\begin{equation}\label{Upsilon_def}
\ket{\Upsilon_{a_1 a_2}}=\frac{1}{\sqrt{d_S}}\sum_{i=1}^{d_S}\ket{i_{a_1}i_{a_2}}.
\end{equation}
can be provided abundantly. They will form the ancillary system $A$ as the state-steering protocol
proceeds (see below). The interaction between $A$ and $E$ is negligible.

\item[(iii)] The state on $\mathcal{H}_S\otimes \mathcal{H}_E$ can be initialised to a fixed
(unknown) pure state $\ket{\Psi_{SE}(0)}$.

\item[(iv)] Any quantum operations can be applied on $SA$ instantaneously.

\item[(v)] State tomography on
SA can be performed at sufficiently high frequency during the protocol
so that we can trace the time evolution of
state $\rho_{SA}(t)$ and its functionals.
\end{itemize}
Although the assumptions (iii) and (iv) lead to (v), we list it here for clarity because it is requisite for our
protocol to work.

Naturally, $d_E$ may be infinitely large in general, but we consider a situation where the system
$S$ effectively interacts with only a finite dimensional subspace $E$ of the \textit{universe}
$E^\prime$. That is, the interaction between $S$ and $E$ is so dominant within the relevant
timescale for describing the dynamics of the system that we can justify this assumption. In other
words, the combined system in $SE$ undergoes a unitary evolution.

For longer timescales, the combined system $SE$ cannot be immune to the effect of interactions
with its surrounding environment $E^\prime$. A state $\rho_{SE}$ is now subject to equilibration
and will tend to some fixed state $\rho_{SE}^{(0)}$. This fact can be used to reset the state
$\rho_{SE}$, albeit unknown, before iterating the protocol. Further, we shall take it for granted
that $\rho_{SE}(0)$ is pure, as assumed in (iii), because we can always purify it by appending an
additional Hilbert space to $E$.

Roughly speaking, we shall present two main results in this paper. One is that, despite the
limited access, it is possible to verify the desired entanglement structure, which is sufficient
for our identification purpose. And the other is that we show by construction the existence of a
protocol to attain the necessary structure in generic situations. In the following subsections, we
give an intuitive description of these two results and that of the tomography for Hamiltonian
identification as well as a remark on quantum control that becomes possible thanks to the
acquisition of those information.

\subsection{Equivalence class and ME condition}\label{subsec:equivalence1}
As described in Sec. \ref{sec:introduction}, we aim at establishing two pairs of maximally
entangled pairs, however, it cannot be directly verified because we are not allowed to access $E$.
Nevertheless, if we could check whether the state on $SA$ satisfies the ME condition, whose details will be explained later in Sec. \ref{sec:pme}, then
the dynamics observable through $SA$ is equivalent to the one that we would see when there were
two maximally entangled pairs as we wanted (cf. Theorem \ref{Theorem:pme_is_mes} in Sec.
\ref{sec:pme_subsec:pme_is_mes}).

\begin{figure}
\includegraphics[scale=0.36]{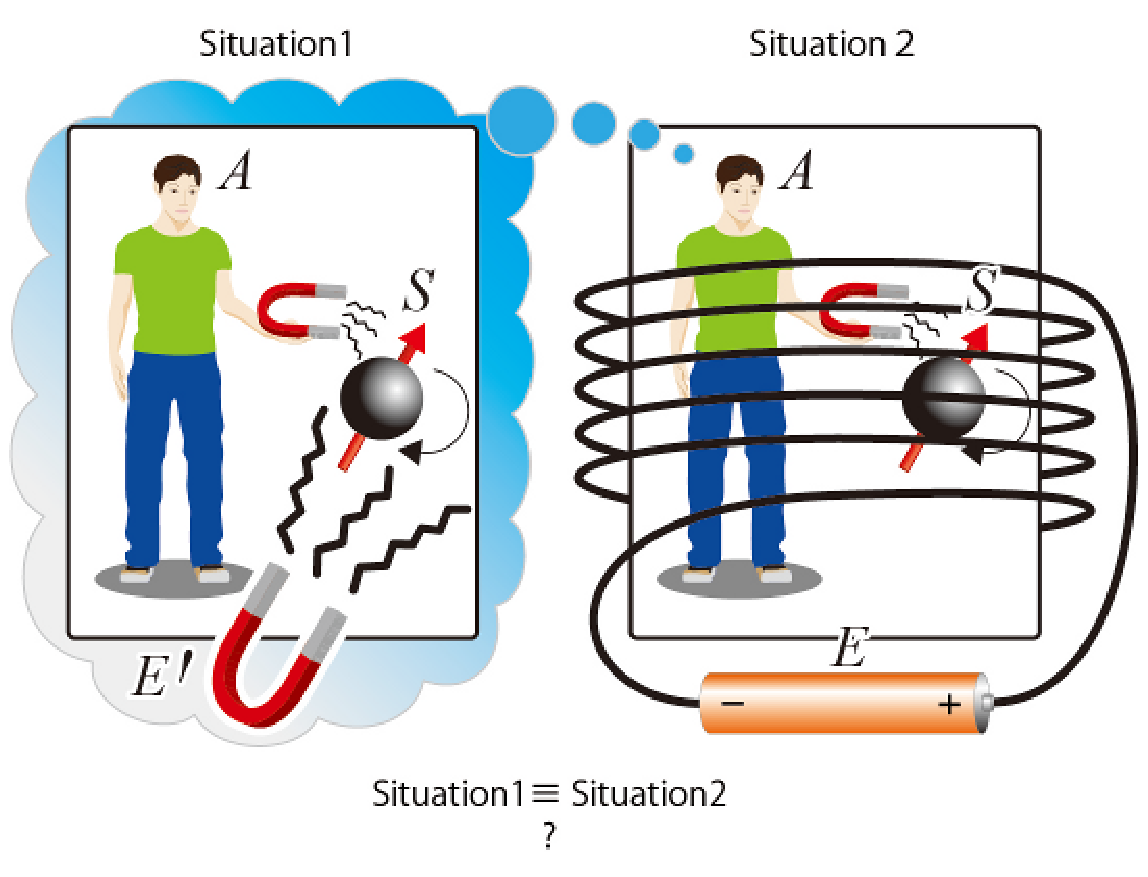}
\caption{An illustration of the physical situations that lead to identical observable dynamics.
Here, the experimenter observes the dynamics a small magnet (spin) shows by varying local control
parameters. Yet, no matter what he controls, there could be multiple possibilities of external
elements, i.e., environment, which would give rise to the same dynamics of the magnet. For
example, he cannot distinguish two situations; the whole laboratory may be in the magnetic field
generated by wire or a permanent magnet. Similarly, in Sec \ref{sec:main_results}, many triples
$\{(d_E, \ket{\Psi_{SEA}},H_{SE})\}$ would lead to indistinguishable dynamics on $SA$.
}\label{fig:equivalence}
\end{figure}

What we mean by equivalent dynamics is as follows, while a mathematically more rigorous treatment
is given in Sec. \ref{subsec:equivalence2}. That is, under the physical situation we consider,
i.e., the one with limited access, there are multiple possibilities of the set $(d_E,
\ket{\Psi_{SEA}}, H_{SE})$ that leads to an identical observable dynamics on $S$ and $A$, no
matter what operations we perform on $SA$. Here, $d_E$, $\ket{\Psi_{SEA}}$ and $H_{SE}$
 are the dimension of $E$, the initial pure state on $SEA$ and the
 Hamiltonian on $SE$. We call this set of three ingredients a
\textit{triple}. Figure \ref{fig:equivalence} illustrates an intuitive picture of the equivalence
between dynamics observed on a subsystem of a larger system.

In the context of system identification, the problem of indistinguishable system models for a
given experimental data set has been studied in the classical setting for a long time
\cite{Ljung_book} and also recently discussed in quantum scenario, in which the entire system is
known to be controllable \cite{Burgarth2012} with a known dimensionality of the system. Our
analysis in this paper is more universal with no extra assumptions, thus the differences between
triples in the same equivalence class look highly nontrivial: even dimensionality can vary within
the class.

With the notion of equivalence class, we can reexpress the content of Theorem
\ref{Theorem:pme_is_mes} in Sec. \ref{sec:pme_subsec:pme_is_mes} as follows. If the state
$\ket{\Psi_{SEA}}$ fulfills the ME condition there exists an equivalent triple in which $A_1$ and
$A_2$ are fully entangled with $S$ and $E$, respectively. Namely,
\begin{equation}\label{two_MES}
\ket{\Psi_{SEA}}=\ket{\Upsilon_{SA_1}}\otimes\ket{\Upsilon_{EA_2}}.
\end{equation}
This fact justifies the use of Eq. (\ref{equivalenceontwosides}) below, or the mirroring effect of
maximally entangled states, for our Hamiltonian tomography, even though the observable $E$ might
be only a subspace that moves around in a larger space.

Since the fulfillment of the ME condition can be checked on $S$ and $A$ only, if any initial state
on $SE$ can be steered to the one that satisfies the ME condition, we can ascertain the
establishment of two maximally entangled pairs. The state-steering protocol, which is depicted in
the next subsection and Sec. \ref{sec:state_steering}, will achieve this task.

For a clear demonstration of the equivalence in terms of the dynamics on $SA$, we have carried out
a numerical simulation of our protocol, taking a four-spin system as an example. In this example,
the single spin of $S$ interacts with each of the three distinct spins of $E$. Since we identify
only one member in the equivalence class that leads to the identical time evolution on $SA$, the
`true' dynamics on $SE$ could be different from what we expect from the estimated
$\tilde{H}_{SE}$. Figure \ref{fig:sim_trace_distance} later in Sec. \ref{sec:numerical} shows how
the difference between dynamics on $SEA$ and that on $SA$ may vary: The observed time evolution on
$SA$ stays the same regardless of the Hamiltonian within the class, i.e., zero trace distance,
while the entire $SEA$ evolves quite differently. For more description, refer to Sec.
\ref{sec:numerical}.

\begin{figure}
\includegraphics[scale=1]{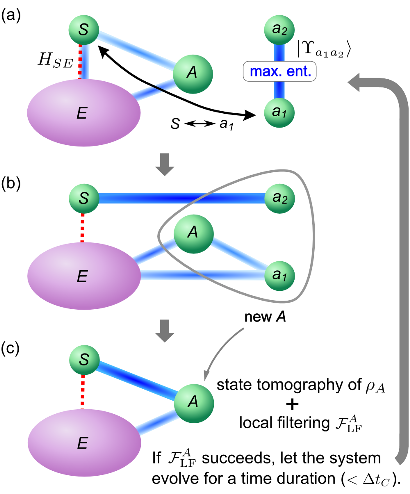}
\caption{The protocol for steering the state on $SEA$ to establish maximal entanglement between
$SE$ and $A$. The thick blue lines and the red dotted line represent entanglement and the
interaction, respectively. The three systems, $S$, $E$, and $A$, are in some entangled state after
foregoing rounds of the protocol, while $A$ is a null space for the first round. In (a),
$\ket{\Upsilon_{a_1 a_2}}$ is provided and the states of $a_1$ and $S$ are swapped to make the
entanglement network look like (b). Relabeling $A$, $a_1$, and $a_2$ as a new $A$ as in (c), we
perform state tomography of $A$ and subsequently a local filtering operation
$\mathcal{F}_\mathrm{LF}^A$ on $A$. If $\mathcal{F}_\mathrm{LF}^A$ succeeds, iterate the
procedure, going back to (a). }\label{fig:protocol}
\end{figure}

\subsection{State-steering protocol}\label{subsec:state-steering}
In order to make use of the \textit{`mirroring effect'} of entanglement, Eq. (\ref{mirroring1}),
for the identification of $H_{SE}$, we first need to steer the state on $SEA$ to establish maximal
entanglement between $SE$ and $A$. Let us describe how the state-steering protocol goes, and
delineate why it works out for our purpose. Figure \ref{fig:protocol} depicts the state-steering
protocol. We start with an initial (fixed, but unknown) state $\rho_{SE}^{(0)}$ and abundant
copies of $\ket{\Upsilon_{a_1 a_2}}$ in Eq. (\ref{Upsilon_def}). At $t=0$ (Step 0),
$\mathcal{H}_{A}$ is a null space, supporting no states. The SWAP operation between $S$ and $a_1$,
which must be fast enough compared with the system dynamics, will be denoted as
$\mathrm{SWAP}_{Sa_1}$. The $C$-th round of the protocol proceeds as follows ($C$ starts from zero at $t=0$):
\begin{itemize}
\item[Step 1:] Apply $\mathrm{SWAP}_{Sa_1}$, where $a_1$ is a subsystem of the newly provided MES,
$\ket{\Upsilon_{a_1 a_2}}$, and then let $A$ incorporate $a_1$ (the former $S$) and $a_2$.

\item[Step 2:] Apply a local filtering operation $\mathcal{F}_\mathrm{LF}^A$ on $\rho_A$ and
increase $C$ by one. If it fails, carry out the whole protocol from the beginning.

\item[Step 3:] Let the $SE$ system evolve for a time duration ($<\Delta t_C$) so that the
functional of $\rho_{SA}$, $\Delta E_{SA}$, which is defined below by Eq. (\ref{def_deltaE_AS}),
increases by $\epsilon_C>0$. See Sec. \ref{sec:state_steering}  as to how we should determine
$\Delta t_C$ and $\epsilon_C$.

\item[Step 4:] Terminate the protocol if $\Delta E_{SA}$ is found to be non-increasing; otherwise,
let the $SE$ system evolve so that $\Delta E_{SA}\ge\epsilon_C$, and go back to Step 1.

\end{itemize}
\textbf{Note that state tomography on $SA$ is performed at sufficiently high frequency during the
protocol so that we can trace the time evolution of state $\rho_{SA}(t)$ and its functionals, such
as $\Delta E_{SA}$ and $\rho_A$.} Intuitively, $\Delta t_C$ and $\epsilon_C$ are set so that we
can complete the steering protocol within a desired time period, which can be made as short as
possible at the expense of success probability.

The local filtering operation on $\rho_A$ is written as $\mathcal{F}_\mathrm{LF}^A
\rho_A=F_\mathrm{LF}\rho_A F_\mathrm{LF}^\dagger$, where
$F_\mathrm{LF}=\sqrt{\lambda_\mathrm{min}\cdot\rho_A^{-1}}$ with $\rho_A^{-1}$ the inverse of
$\rho_A$ on its support and $\lambda_\mathrm{min}$ the smallest nonzero eigenvalue of $\rho_A$.
The success probability of $\mathcal{F}_\mathrm{LF}^A$ is
$\lambda_\mathrm{min}\cdot\mathrm{rank}\rho_A$.

The quantity $\Delta E_{SA}$ we measure in Step 3 is defined as
\begin{equation}\label{def_deltaE_AS}
\Delta E_{SA} := S(\rho_{SA})-S(\rho_A)+\ln d_S,
\end{equation}
where $S(\rho)=-\Tr(\rho\ln\rho)$ is the von Neumann entropy and $\rho_A=\Tr_S\rho_{SA}$. While
what it represents may not be obvious at first sight, $\Delta E_{SA}$ is the change in
entanglement between $SA$ and $E$ due to $\mathrm{SWAP}_{Sa_1}$ in Step 1 of the following round.

We shall see in Sec. \ref{sec:state_steering} that when $\Delta E_{SA}=0$ for any $\Delta t_C$,
there is a subsystem $A_1$ of $A$ that is maximally entangled with $S$. Further, we expect that
the remaining part $A_2$ of $A$ is maximally entangled with $E$ as a result of
$\mathcal{F}_\mathrm{LF}^A$. The saturation of $E_{SA}$, i.e., $\Delta E_{SA}\rightarrow 0$, does
occur because the dimension of $SE$ is finite in our analysis.

\subsection{Tomography for Hamiltonian identification}\label{subsec:ham-identification}
Once two pairs of maximally entangled states are attained, we move on to the Hamiltonian
identification stage. Due to a property of maximally entangled states, for $\ket{\Psi_{SEA}}$ in
Eq. (\ref{two_MES}) and a unitary operator $U_{SE}$, we have
\begin{equation}\label{equivalenceontwosides}
U_{SE}\ket{\Psi_{SEA}}=V_{A} \ket{\Psi_{SEA}},
\end{equation}
where $V_{A}=U_{SE}^T$ acting on $\mathcal{H}_A$, as mentioned above (Eq. (\ref{mirroring1})).
Therefore, the unitary evolution we observe on the ancillary system $A=A_1A_2$ should contain
information about the Hamiltonian $H_{SE}=i/t\ln U_{SE}(t)$.

Naturally, however, simply looking at the state of $A$ does not reveal any information on
$U_{SE}$. As all we can probe is the reduced density operator
$\rho_{SA}(t)=\mathrm{tr}_E\ket{\Psi_{SEA}}\bra{\Psi_{SEA}}$, our task is to find a Hamiltonian
$\tilde{H}_{SE}$ that generates its time evolution $\rho_{SA}(t)$, such that
\begin{equation}\label{timeevolution}
i\frac{\partial}{\partial t}\rho_{SA}=[I_S\otimes \tilde{H}_{SE}^T, \rho_{SA}],
\end{equation}
where $\tilde{H}_{SE}^T$ acts on $\mathcal{H}_A$, despite its notation. (We let a tilde denote the
estimated variable.) The density matrix $\rho_{SA}(t)$ can be Fourier transformed to extract
information about $\tilde{H}_{SE}$ in terms of an orthogonal basis of hermitian operators. The
matrix elements of $\tilde{H}_{SE}$ are then obtained by solving a resulting set of linear
equations. For more details, refer to Sec. \ref{sec:tomography}.

\subsection{Control of $E$ through $S$}\label{subsec:control-of-E}
The Hamiltonian $\tilde{H}_{SE}$ thereby estimated contains all the necessary information to
characterise the observable dynamics, albeit unmodulable per se. What we can control actively is
the system $S$. Thus, the dynamics of the entire system $SE$ is governed by the Hamiltonian
\begin{equation}\label{total_ham}
H(t)=\tilde{H}_{SE}+\sum_n f_n(t)H_S^{(n)},
\end{equation}
where $H_S^{(n)}$ are independent Hamiltonians that act on $\mathcal{H}_S$ and can be modulated by
$f_n(t)$. As we have already identified $\tilde{H}_{SE}$, there is sufficient information to judge
the controllability of the system $SE$ under the Hamiltonian (\ref{total_ham}). A theorem from the
quantum control theory states that the set of realisable unitary operations is generated by
dynamical Lie algebra \cite{Ramakrishna1995,Schirmer2001,D'Alessandro2008}. Dynamical Lie algebra
can be computed by taking all possible (repeated) commutators of operators in Eq.
(\ref{total_ham}), i.e., $i\tilde{H}_{SE}$ and $\{iH_S^{(n)}\}$, and their real linear
combinations.

Therefore, our knowledge of $\tilde{H}_{SE}$ allows the controllable system to encompass not only
the principal system $S$ but also (a part of) the environment $E$. That is, we are now able to
exploit the dynamics inside $E$ for useful quantum operations, such as quantum computing, by
controlling a small system $S$ only. This is the same situation as in refs.
\cite{Lloyd2004,Burgarth2009b,Burgarth2010,Kay2010,Schirmer2008a}, where only a small subsystem is
accessed to control a large system.

\section{Preliminaries and equivalence class of dynamical behaviours}\label{sec:problem_setting}
We will now discuss the problem in a more mathematically rigorous manner. In this section, we give
the definitions and premises of the problem, and define the equivalence class of observable
dynamics under limited access.

\subsection{Problem setting}
We consider a joint system consisting of three parts: the principal system $S$, its surrounding
system (environment) $E$, and an ancillary system $A$, whose Hilbert spaces are denoted as
$\Hi_S$, $\Hi_E$, and $\Hi_A$, respectively. The entire system $SEA$ on $\Hi_S \otimes
\Hi_E\otimes \Hi_A $ is a non-dissipative closed system. The principal system $S$ interacts with
its environment $E$ via Hamiltonian $H_{SE}$, while $E$ does not interact directly with the
ancillary system $A$. Thus, the Hamiltonian of the joint system can be written as $H_{SE} \otimes
I_A$. We are not allowed to access the environmental system $E$ directly, which means that no part
of $E$ can be a subject of direct control or measurement. Meanwhile, we are able to perform any
quantum operations and measurements on the joint system $SA$ instantaneously.

A key assumption we make is that the environmental system $E$ is finite-dimensional, i.e., $d_E :=
\dim E<+\infty$. Those systems under our control, $S$ and $A$ are also finite-dimensional, and
naturally $d_S:=\dim\Hi_S$ and $d_A:=\dim\Hi_A$ are known. We do not assume any prior knowledge of
$d_E$, the interaction Hamiltonian $H_{SE}$, and the state on $\Hi_{SEA}:=\Hi_S\otimes \Hi_E
\otimes \Hi_A$. For most of the discussion in this paper, $d_A$ refers to the
dimension of $A$ after the state-steering protocol (see Sec \ref{sec:main_results} and
\ref{sec:state_steering}) has been completed, unless stated otherwise.

Under these settings, our goal is to obtain as much information as possible about $E$ and the
interaction between $S$ and $E$, namely, $\Hi_E$, $H_{SE}$, and the state on $\Hi_{SEA}$. A
central tool for the information acquisition is quantum state tomography \cite{PR04} of the joint
system $SA$ to determine $\rho_{SA}$ as a function of time. This is possible since the
initialisability of the entire state is assumed so that we can prepare an identical, but not
necessarily known, initial state $\ket{\Psi_{SEA}}$ on  $\Hi_{SEA}$ as many times as necessary.
Pragmatically, such a state initialisation can be achieved by waiting for the equilibration of the
state, which is caused by the interaction with a larger environmental system that surrounds $E'$
\cite{OPENQUANTUM,NIELSEN}. The timescale for such an equilibration is much longer than the one
within which the entire system of $SEA$ can be considered closed. The initial time $t_0$ is
defined as the time when the state initialisation is completed.

Since we have prior information about $d_S$ and $d_A$, our system dynamics can be characterised by
$d_E$, a state on $\Hi_{SEA}$ at the time $t_0$, and the Hamiltonian $H_{SE}$. Note that at $t_0$,
we can always assume the initial state on $\Hi_{SEA}$ is pure. This is because when the state on
$\Hi_{SEA}$ is mixed, we can append an extra Hilbert space $\Hi_{F}$ to the system so that
the state on $\Hi_{SEA}\otimes\Hi_{F}$ is pure \cite{NIELSEN,HAYASHI}. Then, we simply
redefine $\Hi_E \otimes \Hi_{F}$ as $\Hi_E$, and $H_{SE}\otimes I_{F}$ as a new
Hamiltonian $H_{SE}$ to restart the whole discussion. Therefore, we need a set of three elements
$(d_E, \ket{\Psi_{SEA}}, H_{SE})$, which we shall call a \textit{triple}, to characterise the
behaviour of our system under the effect of environment between the times $t_0$ and $t_\infty$
$(t_0 < t_\infty)$. Although we choose $t_\infty =+ \infty$, which is theoretically natural, since
we practically perform an experiment within a finite time length, the consideration of a finite
$t_\infty$ is also useful as we will see later.

\subsection{The equivalence due to indistinguishable dynamics}\label{subsec:equivalence2}
As mentioned above, our primary goal is to identify the triple $(d_E, \ket{\Psi_{SEA}}, H_{SE})$.
Yet, it is an impossible task to completely specify the triple when our access is limited to only
$S$ and $A$, no extra assumptions are given. What if there are more than one possible triple?
Similarly to the cases studied in the past \cite{Ljung_book,Burgarth2012}, for our task of
identifying the system $E$, it turns out that even if there were multiple possibilities of triples
that give rise to the same dynamical behaviour on $SA$, the difference between them would not lead
to distinct outcomes of quantum control of $E$ through $S$. In other words, if there were two
indistinguishable environmental systems characterised by $(d_E, \ket{\Psi_{SEA}}, H_{SE})$ and
$(\tilde{d}_E, \ket{\tilde{\Psi}_{SEA}}, \tilde{H}_{SE})$ in the time period $[t_0, t_\infty)$,
the results of any quantum computation that utilises $E$ as a resource would be independent of
whether the \textit{true} environment was either of them. Therefore, for the acquisition of
information on the environmental system $\Hi_E$ toward the exploitation of $E$ as a (partial)
resource for quantum computing, it suffices to determine the equivalence class on the set of all
triples through all possible sequences of quantum operations on $\Hi_S \otimes \Hi_A$.

Before giving a rigorous definition of the equivalence class of triples, let us first specify all
operations we can apply on the system. First, we do not consider operations that are applied
continuously in time. Thus, what we consider to be applicable is a sequence of instantaneous
quantum operations \cite{NIELSEN,HAYASHI,HOLEVO1972,KRAUS}, $\{ \Gamma_i\}_{i=1}^n$ at time $t_i$,
where $n < +\infty$ and $t_i < t_j$ for all $i<j$. Second, a quantum operation $\Gamma_i$ can be
non-deterministic, i.e., trace non-increasing, because we can always post-select the measurement
results. Third, we are allowed to append and remove finite-dimensional ancillary systems, which
means that $\Gamma_i$ is a quantum operation on $\B\left( \Hi_{A_{i-1}} \otimes \Hi_S \right)$ to
$\B\left( \Hi_{A_{i}} \otimes\Hi_S \right)$, where $\B\left(\Hi \right)$ is a linear space of all
(bounded) linear operators on $\Hi$, $\Hi_{A_{i-1}} \neq \Hi_{A_i}$ in general, $\dim \Hi_{A_i} <
+\infty$, and $\Hi_{A_0} := \Hi_A$.

Hence, our definition of the equivalence class is as follows:

\begin{Definition}\label{equiv_env}
\textbf{(The equivalence between triples)} A triple $(d_E, \ket{\Psi_{SEA}}, H_{SE})$ is said to
be equivalent to another triple, $(\tilde{d}_E, \ket{\tilde{\Psi}_{SEA}},\tilde{H}_{SE})$, in
$[t_0, t_\infty)$, if they satisfy
\begin{align}
& \Tr _E \left( \prod_{i=1}^n   \left(\Gamma _i \otimes \mathcal{I}_E \right) \circ
\left( \mathcal{I}_A \otimes \mathcal{U}_{SE}^{(i)} \right)
 P(\ket{\Psi_{SEA}}) \right)\nonumber \\
=& \Tr _E \left( \prod_{i=1}^n   \left(\Gamma _i \otimes \mathcal{I}_E \right) \circ \left(
\mathcal{I}_A \otimes \tilde{\mathcal{U}}_{SE}^{(i)} \right) P(\ket{\tilde{\Psi}_{SEA}}) \right)
\label{A2_eq_definition_of_equivalence}
\end{align}
for all $n \in \mathbb{N}$ and all sequences of completely positive trace non-increasing maps $\{
\Gamma _i \}_{i=1}^n$ \cite{NIELSEN,HAYASHI,HOLEVO1972,KRAUS}. Each $\Gamma_i$ is a map from
$\B\left( \Hi_{A_{i-1}} \otimes  \Hi_S \right)$ to $\B \left( \Hi_{A_{i}} \otimes  \Hi_S\right)$
and performed at time $t_i \,(i\in\{1,2,...,n\})$. We will denote the equivalence between triples
as $(d_E, \ket{\Psi_{SEA}}, H_{SE}) \equiv (\tilde{d}_E, \ket{\tilde{\Psi}_{SEA}},
\tilde{H}_{SE})$.
\end{Definition}
In Eq.~(\ref{A2_eq_definition_of_equivalence}), $\mathcal{I}_A$ and $\mathcal{I}_E$ are identity
(super)operators on $\Hi_A$ and $\Hi_E$, respectively. Also, $P(\ket{\Psi})$ stands for
$P(\ket{\Psi}) := \ket{\Psi}\bra{\Psi}$. $\mathcal{U}^{(i)}_{SE}$ is given as
\[
\mathcal{U}^{(i)}_{SE}(\rho) := \exp \left(-i H_{SE} (t_i-t_{i-1}) \right) \rho \exp \left(i
H_{SE} (t_i-t_{i-1}) \right),
\]
and $\tilde{\mathcal{U}}^{(i)}_{SE}(\rho)$ is defined similarly with $\tilde{H}_{SE}$ instead of
$H_{SE}$. We can easily see that the relation ``$\equiv$" is reflective, symmetric, and
transitive. Thus, it is an equivalence relation in the mathematical sense \cite{KOLMOGOROV}, and a
set of all the triples can be decomposed into equivalence classes accordingly.

Our definition of equivalence here differs from the one in \cite{Burgarth2012} in that ours
includes the possibility of appending an arbitrarily large ancillary system. Also, in Def.
\ref{equiv_env} above, the controllability of the system of interest is not assumed. Such a
consideration is important especially when we can utilise the joint system $SE$ as a part of a
larger quantum network, rather than an isolated quantum computer.

\begin{figure}
\includegraphics[scale=1]{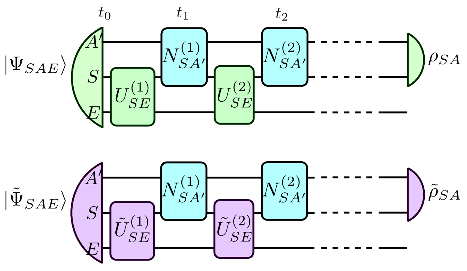}
\caption{(Color online) The definition of equivalence. Suppose there are two situations that are
characterised by triples $(d_E, \ket{\Psi_{SEA}}, H_{SE})$ and $(\tilde{d}_E,
\ket{\tilde{\Psi}_{SEA}}, \tilde{H}_{SE})$, respectively. If the same sequence of quantum
operations $\{N_{SA^\prime}^{(j)}\,(j=1,2,...)\}$, each of which is applied instantaneously on
$SA^\prime$ at time $t_j$, leads to the same state on $SA^\prime$, i.e., $\rho_{SA^\prime}=\tilde
\rho_{SA^\prime}$, we call the two situations are equivalent, denoting $(d_E, \ket{\Psi_{SEA}},
H_{SE})\equiv (\tilde{d}_E, \ket{\tilde{\Psi}_{SEA}}, \tilde{H}_{SE}).$ Here, the Hilbert space
$\mathcal{H}_{A^\prime}$ includes $\mathcal{H}_A$.}\label{fig:EquivalenceDiagram}
\end{figure}

Let us slightly simplify the definition of the above equivalence relation for the following
discussion. Here, we define a Hilbert space $A^\prime$ that includes all $A_i$ as its subspace.
\begin{Lemma}\label{lemma:sec_problem_setting}
A triple $(d_E, \ket{\Psi_{SEA}}, H_{SE})$ is equivalent to anther one $(\tilde{d}_E,
\ket{\tilde{\Psi}_{SEA}}, \tilde{H}_{SE})$ in $[t_0, t_\infty)$, if and only if they satisfy
\begin{align}
& \Tr _E P\left(  \left( \prod_{i=1}^n  N_{SA^\prime}^{(i)}  \cdot U_{SE}^{(i)} \right)
 \ket{\Psi_{SEA}}  \right) \nonumber \\
=& \Tr _E P\left( \left( \prod_{i=1}^n   N_{SA^\prime}^{(i)} \cdot  \tilde{U}_{SE}^{(i)} \right)
 \ket{\tilde{\Psi}_{SEA}}  \right)
 \label{A2_eq_definition_of_equivalence_2}
\end{align}
for all $n \in \mathbb{N}$, all sequences of real numbers $\{t_i \}_{i=1}^n$ with $t_i<t_j$ for
$i<j$, all finite dimensional Hilbert spaces $\Hi_{A^\prime}$ which includes $\Hi_A$ as a subspace
($\Hi_A \subset \Hi_{A^\prime} $), and all sequences of operators $\{ N_{SA^\prime}^{(i)}
\}_{i=1}^n$ on $\Hi_{A^\prime} \otimes \Hi_S$.
\end{Lemma}
A quantum circuit representation of Eq. (\ref{A2_eq_definition_of_equivalence_2}) is shown in Fig.
\ref{fig:EquivalenceDiagram}, where LHS and RHS are denoted as $\rho_{SA}$ and $\tilde \rho_{SA}$
for short. In Eq.~(\ref{A2_eq_definition_of_equivalence_2}), $U_{SE}^{(i)} := \exp \left[ -i
H_{SE}(t_i-t_{i-1}) \right]$, and $I_E$ and $I_{A^\prime}$ are omitted. We shall not write
identity operators explicitly throughout the paper when there is no risk of confusion.
\\
{\bf (Proof)} \\
\quad In order to prove the ``{\it only if}" part, since $\Gamma_i(\rho) := N_{SA^\prime}^{(i)}
\rho N_{SA^\prime}^{(i)\dagger}$ is a trace non-increasing CP map, we simply need to define
$\Hi_{A^\prime}$ as $\Hi_{A^\prime} :=\Hi_{A_i}$. For the ``{\it if}" part, we define
$\Hi_{A^\prime} := \bigoplus_{i=0}^n \Hi_{A_i}$. Then, as any trace non-increasing CP map can be
expressed as a sum of the terms, each of which has the form of Eq.
(\ref{A2_eq_definition_of_equivalence_2}), the linearity of the partial trace $\Tr_E$ guarantees
the statement of the lemma. $\hfill \blacksquare$

\section{The maximal entanglement condition and the equivalence between triples}\label{sec:pme}
We now introduce a condition that is of crucial importance for our analysis. We shall refer to it
as the maximal entanglement (ME) condition. It will be shown that when our system satisfies this
condition, the natural time evolution of the system $SA$ completely determines the equivalence
class (the subsection \ref{subsec:MEcondition}). By natural time evolution, we mean the evolution
of the system without active operations on it, namely, the evolution that is driven only by the
system Hamiltonian, which is $H_{SE}$ in our case. Then, we prove that when the system satisfies
the ME condition, there exists an equivalent triple, in which $E$ which is maximally entangled
\cite{NIELSEN,WERNER1989,HORODECKI2009} with (a subset of) $A$ (the subsection
\ref{sec:pme_subsec:pme_is_mes}). We note that although our main focus is on a finite-dimensional
$\Hi_E$ in this paper, some of the theorems and lemmas in this section are valid even for
infinite-dimensional $\Hi_E$, as long as $d_A$ and $d_S$ are finite.

\subsection{The maximal entanglement condition} \label{subsec:MEcondition}
The condition for the equivalence of triples still appears quite complicated even in the form of
 Lemma \ref{lemma:sec_problem_setting}. In this subsection, we prove that the condition for
the equivalence reduces to merely the indistinguishability of the natural time evolution of the
system $SA$, when the condition defined in the following is satisfied:
\begin{Definition}\label{Def:MECondition}
\textbf{(The maximal entanglement condition)} The maximal entanglement (ME) condition is said to
be satisfied by a triple $(d_E, \ket{\Psi_{SEA}}, H_{SE})$ in $[t_0, t_\infty)$, if for all $t\in
[t_0,t_\infty)$, there exist Hilbert spaces $\Hi_{A_1}(t)$ and $\Hi_{A_2}(t)$ such that $\Hi_A =
\Hi_{A_1}(t) \otimes \Hi_{A_2}(t)$, $\dim \Hi_S = \dim \Hi_{A_1}(t)$, and
\begin{equation}\label{eq:def_pme}
\Tr _E P\left( \ket{\Psi_{SEA}(t)} \right)= P\left( \ket{\Upsilon_{SA_1}(t)}\right) \otimes
\rho_{A_2}(t),
\end{equation}
where $\ket{\Upsilon_{SA_1}(t)}$ is a maximally entangled state on $\Hi_S \otimes \Hi_{A_1}(t)$
and a state $\rho_{A_2}(t)$ is a projector onto $\Hi_{A_2}(t)$ (up to a proportionality constant).
\end{Definition}
Throughout this paper, $\ket{\Upsilon_{XY}}$ denotes a state that is maximally entangled
\textit{fully} on the space $\mathcal{H}_X\otimes\mathcal{H}_Y$ specified by the subscripts. That
is, $\ket{\Upsilon_{XY}}=d_X^{-1/2}\sum_{i=1}^{d_X}\ket{i_X}\ket{i_Y}$, where $\{\ket{i_X}\}$ and
$\{\ket{i_Y}\}$ are arbitrary orthonormal bases of $\mathcal{H}_X$ and $\mathcal{H}_Y$, and
$d_X=\mathrm{dim} \mathcal{H}_X (=d_Y)$.

In the above definition, $\ket{\Psi_{SEA}(t)}$ is the entire state at time $t$, i.e.,
\begin{equation}\label{eq:def_psi_t}
\ket{\Psi_{SEA}(t)} := \exp(-iH_{SE}(t-t_0))\ket{\Psi_{SEA}}.
\end{equation}
Here, we note that the Hilbert spaces $\Hi_{A_1}(t)$ and $\Hi_{A_2}(t)$ may vary inside $\Hi_A$ as
the time evolution of $SE$ (due to $H_{SE}$) would be reflected in $A_1$ and $A_2$ through
entanglement.

Equation (\ref{eq:def_pme}) implies the existence of a pure state $\ket{\Phi_{EA_2}(t)}$ on
$\Hi_E\otimes\Hi_{A_2}(t)$ such that
\begin{equation}\label{eq:def_pme2}
\ket{\Psi_{SEA}(t)}=\ket{\Upsilon_{SA_1}(t)}\otimes \ket{\Phi_{EA_2}(t)},
\end{equation}
where $\ket{\Phi_{EA_2}(t)}$ is a maximally entangled state (MES) on a subspace of
$\Hi_E\otimes\Hi_{A_2}(t)$. Yet the rank of the reduced density matrix $\rho_E$ may be smaller
than the dimension of the system $E$: ${\rm rank}\rho_E < d_E$. It turns out, however, it is
possible to choose a triple $(\tilde{d}_E, \ket{\tilde{\Psi}_{SEA}},\tilde{H}_{SE})$ (equivalent
to $(d_E, \ket{\Psi_{SEA}}, H_{SE})$) so that $\ket{\tilde{\Phi}_{EA_2}(t)}$ can be expressed as
$\ket{\Upsilon_{EA_2}}=\tilde{d}_E^{-1/2}\sum_{i=1}^{\tilde{d}_E}\ket{i_{A_2}}\ket{i_E}$, i.e., a
state that is not only maximally entangled but also satisfies ${\rm rank}\tilde{\rho}_E =
\tilde{d}_E$ (see Theorem \ref{Theorem:pme_is_mes} and Corollary
\ref{sec:pme_subsec:tomography_corollary_1}), where $\tilde{\rho}_E=\Tr_{A_2}
\ket{\Upsilon_{EA_2}}\bra{\Upsilon_{EA_2}}$. When the ME condition is found to be satisfied, we
can take it for granted that the whole $E$ is maximally entangled with $A_2\subset A$, despite the
inaccessibility of $E$.

An observation is that the fulfillment of the ME condition can be tested through tomography of the
state on $SA$ only. This fact leads to a lemma:
\begin{Lemma}\label{ME:lemma_1}
Suppose $(d_E, \ket{\Psi_{SEA}}, H_{SE})$ satisfies the ME condition in $[t_0, t_\infty)$, and
$(d_E,  \ket{\Psi_{SEA}}, H_{SE}) \equiv (\tilde{d}_E, \ket{\tilde{\Psi}_{SEA}}, \tilde{H}_{SE})$
in $[t_0, t_\infty)$. Then, $(\tilde{d}_E, \ket{\tilde{\Psi}_{SEA}}, \tilde{H}_{SE})$ also
satisfies the ME condition in $[t_0, t_\infty)$.
\end{Lemma}
{\bf (Proof)} Due to the definition of the equivalence, if $(d_E, \ket{\Psi_{SEA}}, H_{SE}) \equiv
(\tilde{d}_E, \ket{\tilde{\Psi}_{SEA}}, \tilde{H}_{SE})$ in $[t_0, t_\infty)$, we have, for all $t
\in [t_0, t_\infty)$,
\begin{equation}\label{eq:ME_lemma_1}
\Tr_E P\left (\ket{\Psi_{SEA}(t)} \right )=\Tr_E P \left(\ket{\tilde{\Psi}_{SEA}(t)} \right),
\end{equation}
where $\ket{\Psi_{SEA}(t)}$ is given in Eq. (\ref{eq:def_psi_t}) and $\ket{\tilde{\Psi}_{SEA}(t)}$
is defined similarly. Then, since the fulfillment of the ME condition only depends on the reduced
density operator on $\Hi_S \otimes\Hi_A$ during $[t_0,t_\infty)$, Eq.~(\ref{eq:ME_lemma_1})
guarantees the statement of the lemma.
\hfill $\blacksquare$ \\
Lemma \ref{ME:lemma_1} implies that the ME condition can also be considered as a property of
equivalence class of triples.

Let us now present a theorem, which claims that if two triples give rise to an identical (natural)
time evolution on $SA$ and if one of them satisfies the ME condition, then those two triples are
equivalent.

\begin{Theorem}\label{MEtheorem}
Suppose $(d_E, \ket{\Psi_{SEA}}, H_{SE})$ satisfies the ME condition in $[t_0, t_\infty)$. Then,
$(d_E, \ket{\Psi_{SEA}}, H_{SE})\equiv (\tilde{d}_E, \ket{\tilde{\Psi}_{SEA}}, \tilde{H}_{SE})$ in
$[t_0, t_\infty)$, if and only if they satisfy
\begin{equation}\label{eq:MEtheorem_1}
 \Tr _E P \left( \ket{\Psi_{SEA}(t)}\right) = \Tr _E P \left( \ket{\tilde{\Psi}_{SEA}(t)}\right)
\end{equation}
for all $t\in [t_0, t_\infty)$.
\end{Theorem}
Thus, when the ME condition is satisfied, an identical time evolution on $SA$ is sufficient to
certify the equivalence, that is, we do not need to consider all possible sequences of quantum
operations $\{\Gamma_i\}$. Towards the proof of the theorem, we show two lemmas.

\begin{figure}
\includegraphics[scale=1]{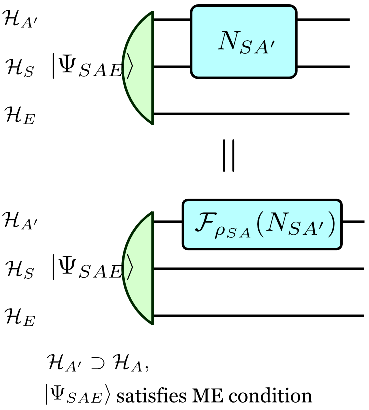}
\caption{(Color online) Schematic illustration of the relation in Eq. (\ref{eq:ME_lemma3_2}). The
Hilbert space $\mathcal{H}_{A^\prime}$ includes $\mathcal{H}_A=\mathcal{H}_{A_1}\otimes
\mathcal{H}_{A_2}$, and the state $\ket{\Psi_{SEA}}$ satisfies the ME condition. No matter what
operation $N_{SA^\prime}$ is performed on $SA^\prime$, the same effect can be realised by an
operation $\mathcal{F}_{\rho_{SA}}(N_{SA^\prime})$ applied solely on $A^\prime$. $\mathcal{F}$ is
a functional of $N_{SA^\prime}$ and the state $\rho_{SA}$.}\label{fig:ProofOfLemma3}
\end{figure}

\begin{Lemma}\label{ME:lemma3}
Suppose a state $\ket{\Psi_{SEA}}$ on the Hilbert space $\Hi_S \otimes \Hi_E \otimes \Hi_{A_1}
\otimes \Hi_{A_2},$ where $d_S= \dim \Hi _S = \dim \Hi_{A_1}$, and $\ket{\Psi_{SEA}}$ can be
written as
\begin{equation}\label{eq:def_pme3}
 \ket{\Psi_{SEA}}=\ket{\Upsilon_{SA_1}}\otimes \ket{\Phi_{EA_2}},
\end{equation}
where $\ket{\Upsilon_{SA_1}}$ is a maximally entangled state, and $\Tr_E P\left(
\ket{\Phi_{EA_2}}\right)$ is a projector up to a proportionality
 constant. Then, for all finite-dimensional Hilbert spaces
 $\Hi_{A^\prime}$ that include $\Hi_{A_1} \otimes \Hi_{A_2}$ as a
subspace, and for all operators $N_{SA^\prime}$ on $\Hi_S \otimes \Hi_{A^\prime}$,
\begin{enumerate}
 \item ${\rm supp} \Tr _{A^\prime}P\left( N_{SA^\prime} \ket{\Psi_{SEA}} \right) \subset
       {\rm supp} \Tr _A P\left( \ket{\Psi_{SEA}}\right)$

 \item For a given state $\rho_{SA}:= \Tr _E P \left(\ket{\Psi_{SEA}}\right)$,
there exists a linear map $\mathcal{F}_{\rho_{SA}}$ from $\B
    \left(\Hi_S \otimes \Hi_{A^\prime}\right)$ to $\B\left( \Hi_{A^\prime} \right)$
such that
\begin{equation}\label{eq:ME_lemma3_2}
 \mathcal{F}_{\rho_{SA}}\left(N_{SA^\prime}\right)\otimes
  I_{SE}\ket{\Psi_{SEA}}=N_{SA^\prime}\otimes I_E\ket{\Psi_{SEA}},
\end{equation}
for all $N_{SA^\prime} \in \B\left(\Hi_S\otimes\Hi_{A^\prime}\right)$.
\end{enumerate}
\end{Lemma}
Here, we note that the same $\mathcal{F}_{\rho_{SA}}$ satisfies Eq. (\ref{eq:ME_lemma3_2}) for all
states $\ket{\Psi_{SEA}}$ having the common reduced density matrix $\rho_{SA}= \Tr _EP\left(
\ket{\Psi_{SEA}}\right)$. Figure \ref{fig:ProofOfLemma3} depicts the equivalence relation of Eq.
(\ref{eq:ME_lemma3_2}).

{\bf (Proof)} \\
The proof of the first statement proceeds as follows. We write $\ket{\Upsilon_{SA_1}}$ and
$\ket{\Phi_{EA_2}}$ as
\begin{align*}
 \ket{\Upsilon_{SA_1}} &= \frac{1}{\sqrt{d_S}}\sum_{i=1}^{d_S} \ket{e_i}
 \otimes \ket{i}, \\
\ket{\Phi_{EA_2}} &= \frac{1}{\sqrt{r}}\sum_{j=1}^{r} \ket{f_j}
 \otimes \ket{j},
\end{align*}
where $\left\{ \ket{e_i} \right\}_{i=1}^{d_S}$, $\left\{ \ket{f_j} \right\}_{j=1}^{d_E}$, $\left\{
\ket{i} \right\}_{i=1}^{d_S}$, and $\left\{ \ket{j} \right\}_{j=1}^{\dim \Hi_{A_2}}$ are
orthonormal bases of $\Hi_S, \Hi_E, \Hi_{A_1}$, and $\Hi_{A_2}$, respectively, and $r$ is the
Schmidt rank \cite{NIELSEN} of $\ket{\Phi_{EA_2}}$. Defining $n_{i,j,k}$ to be
$\sum_{i'=1}^{d_s}{}_S
\bra{e_i}_{A'}\bra{\eta_k}N_{SA'}\ket{e_{i'}}_S\ket{i'}_{A_1}\ket{j}_{A_2}/\sqrt{d_Sr}$, we see
the following equality:
\begin{equation}\label{eq:supportproof}
 N_{SA^\prime}  \ket{\Psi_{SEA}}
 =
 \sum
  _{i=1}^{d_S}
 \sum
  _{j=1}^{r}
 \sum
  _{k=1}^{dim A'}
n_{i,j,k} \ket{e_i}\ket{f_j}\ket{\eta_k}_{A^\prime},
\end{equation}
where $\left\{ \ket{\eta_k} \right\}_{\alpha=1}^{\dim A^\prime}$  is an orthonormal basis of
$\Hi_{A^\prime}$. As the basis vectors of $\Hi_E$ in the summation are restricted to
$\{\ket{f_j}\}$ with $1\le j\le r$, Eq. (\ref{eq:supportproof}) implies ${\rm
supp}\Tr_{A'}P(N_{SA'}\ket{\Psi_{SEA}})\subset
\mathrm{span}\left\{\ket{e_i}\ket{f_j}\right\}_{i=1,j=1}^{d_S,\ r} $, hence statement 1.

We now move on to the proof of the second statement. Suppose $\left\{ \ket{\eta_\alpha}
\right\}_{\alpha=1}^{d_{A^\prime}}$ is an orthonormal basis of $\Hi_{A^\prime}$, where
$d_{A^\prime}=\dim \Hi_{A^\prime}$. There is a natural linear isomorphism, $g$, from
$\Hi_{A^\prime} \otimes {\rm supp} \Tr _A P\left( \ket{\Psi _{SEA}}\right)$ to $\B \left( {\rm
supp} \Tr_A P\left( \ket{\Psi _{SEA}}\right), \Hi_{A^\prime} \right)$ ($\B ( \Hi, \mathcal{K})$ is
a space of all linear operators from $\Hi$ to $\mathcal{K}$). $g$ is defined by the correspondence
of their bases:
\begin{equation}
 g: \ket{e_i} \ket{f_j} \ket{\eta_\alpha}\mapsto \ket{\eta_\alpha}
  \bra{e_i} \bra{f_j} \quad (\forall \alpha, i,j).
\end{equation}
Then, we define a linear map $\mathcal{F}_{\rho_{SA}}$  as
\begin{equation}\label{eq:def_F_proof_th2}
 \mathcal{F}_{\rho_{SA}}(N_{SA^\prime}) := \sqrt{d_S r}g\left( N_{SA^\prime} \ket{\Psi_{SEA}}\right)V,
\end{equation}
where $N_{SA^\prime} \in \B\left(\Hi_S\otimes\Hi_{A^\prime}\right)$, and $V$ is a partial isometry
defined as
\[
V := \sum_{i=1}^{d_S} \sum_{j=1}^r \ket{e_i}_{S}\ket{f_j}_E\bra{i}_{A_1}\bra{j}_{A_2}.
\]
We note that $\mathcal{F}_{\rho_{SA}}$ is well defined thanks to statement 1. Then, it is
straightforward to see that $\mathcal{F}_{\rho_{SA}}$ satisfies Eq.~(\ref{eq:ME_lemma3_2}).

It is possible to show that the linear map $\mathcal{F}_{\rho_{SA}}$ only depends on the reduced
density matrix $\rho_{SA}$ of the state $\ket{\Psi}$; more specifically, its effect does not
depend on the Schmidt bases $\{\ket{e_i}\}_{i=1}^{d_S}$ and $\{ \ket{f_i} \}_{i=1}^{d_E}$.
Although $g$ and $V$ do depend on $\{\ket{e_i}\}_{i=1}^{d_S}$ and $\{ \ket{f_i} \}_{i=1}^{d_E}$,
their dependence cancels out in Eq. (\ref{eq:def_F_proof_th2}). As a result, any specific choice
of these bases does not affect the action of $\mathcal{F}_{\rho_{SA}}$. We can also verify this
property of $\mathcal{F}_{\rho_{SA}}$ from the fact that equation
$\mathcal{F}_{\rho_{SA}}\left(N_{SA^\prime}\right)U_E\ket{\Psi_{SEA}}=N_{SA^\prime}U_E\ket{\Psi_{SEA}}$
holds for all unitary operators $U_E$ on $\Hi_E$. Therefore, statement 2 holds. \hfill
$\blacksquare$

\begin{Lemma}\label{ME:lemma4}
Suppose $(d_E, \ket{\Psi_{SEA}}, H_{SE})$ satisfies the ME condition in $[t_0, t_\infty)$. Then,
for all instances $\left\{ t_i \right\}_{i=1}^n$, where $t_0 < t_i < t_j < t_\infty$ for $i<j$,
all finite dimensional Hilbert spaces $\Hi_{A^\prime}\supset \Hi_{A}$, and all sets of linear
operators $\left\{ N_{SA^\prime} ^{(i)} \right\}_{i=1}^n$ on $\Hi_S\otimes\Hi_{A^\prime}$, the
following equation holds:
\begin{align}
& \prod_{j=1}^n \left( N_{SA^\prime}^{(j)} U_{SE}^{(j)}\right)\ket{\Psi_{SEA}}
 \nonumber \\
=& \left(\left( \mathcal{F}_{\rho_{SA}(t_n)} \cdot \prod _{j=2}^n N_{SA^\prime}^{(j)}
\mathcal{F}_{\rho_{SA}(t_{j-1})}
\right) \left( N_{SA^\prime}^{(1)} \right) \right) \otimes I_{SE} \nonumber \\
& \qquad \ket{\Psi_{SEA}(t_n)}, \label{eq:psi_j=1_left_n_SA}
\end{align}
where $N_{SA^\prime}\mathcal{F}_{\rho}$ is a linear map on $\B \left(\Hi_S\otimes\Hi_{A^\prime}
\right)$ defined as $N_{SA^\prime}\mathcal{F}_{\rho}(M_{SA})= N_{SA^\prime}\left(
\mathcal{F}_{\rho}(M_{SA^\prime}) \otimes I_S \right)$ for $M_{SA^\prime} \in \B\left(
\Hi_S\otimes\Hi_{A^\prime}\right)$.
\end{Lemma}
{\bf (Proof)} \\
By repeatedly applying Lemma \ref{ME:lemma3}, we have:
\begin{eqnarray}
& & \prod_{j=1}^n \left( N_{SA^\prime}^{(j)} U_{SE}^{(j)}\right)\ket{\Psi_{SEA}}
 \nonumber \\
&=& \prod_{j=2}^n \left( N_{SA^\prime}^{(j)} U_{SE}^{(j)}\right)
 N_{SA^\prime}^{(1)} \ket{\Psi_{SEA}(t_1)}
 \nonumber \\
&=& \prod_{j=2}^n \left( N_{SA^\prime}^{(j)} U_{SE}^{(j)}\right)
 \mathcal{F}_{\rho_{SA}(t_1)} \left( N_{SA^\prime}^{(1)}\right) \ket{\Psi_{SEA}(t_1)}
 \nonumber \\
&=& \prod_{j=3}^n \left( N_{SA^\prime}^{(j)} U_{SE}^{(j)}\right)
 N_{SA^\prime}^{(2)}\mathcal{F}_{\rho_{SA}(t_1)} \left( N_{SA^\prime}^{(1)}\right) \ket{\Psi_{SEA}(t_2)}
 \nonumber \\
&=& \prod_{j=3}^n \left( N_{SA^\prime}^{(j)} U_{SE}^{(j)}\right) \mathcal{F}_{\rho_{SA}(t_2)}
\left(
 N_{SA^\prime}^{(2)}\mathcal{F}_{\rho_{SA}(t_1)} \left( N_{SA^\prime}^{(1)}\right)
 \right) \nonumber \\
& & \times\ket{\Psi_{SEA}(t_2)}
 \nonumber \\
&=& \left(\left( \mathcal{F}_{\rho_{SA}(t_n)} \cdot \prod _{j=2}^n
 N_{SA^\prime}^{(j)} \mathcal{F}_{\rho_{SA}(t_{j-1})}
\right) \left( N_{SA^\prime}^{(1)} \right) \right) \otimes I_{SE}  \nonumber \\
& & \times\ket{\Psi_{SEA}(t_n)}.
\end{eqnarray}
This proves Eq. (\ref{eq:psi_j=1_left_n_SA}).
\hfill $\blacksquare$\\

We are now ready to prove Theorem \ref{MEtheorem}.\\
{\bf (Proof of Theorem \ref{MEtheorem})} \\
The ``{\it only if}" part is trivial from the definition. We therefore prove the ``{\it if}" part
of the statement. Suppose $(d_E, \ket{\Psi_{SEA}}, H_{SE})$ satisfies the ME condition and $(d_E,
\ket{\Psi_{SEA}}, H_{SE})$ and  $(\tilde{d}_E, \ket{\tilde{\Psi}_{SEA}}, \tilde{H}_{SE})$ satisfy
Eq. (\ref{eq:MEtheorem_1}). Then, for all instances $\left\{ t_i \right\}_{i=1}^n$, where $t_0 <
t_i < t_j < t_\infty$ for $i<j$, all finite-dimensional Hilbert spaces $\Hi_{A^\prime}\supset
\Hi_{A}$ and all sets of linear operators $\left\{ N_{SA^\prime} ^{(i)} \right\}_{i=1}^n$ on
$\Hi_S\otimes\Hi_{A^\prime}$, we derive the following equations:
\begin{eqnarray}\label{ProofMETheorem}
& & \Tr _E P \left( \prod_{j=1}^n \left( N_{SA^\prime}^{(j)} U_{SE}^{(j)}\right)
 \ket{\Psi_{SEA}}\right) \nonumber \\
&=& \left(\left( \mathcal{F}_{\rho_{SA}(t_n)} \cdot \prod
 _{j=2}^n N_{SA^\prime}^{(j)} \mathcal{F}_{\rho_{SA}(t_{j-1})}
\right) \left( N_{SA^\prime}^{(1)} \right) \right) \otimes I_{S} \cdot \nonumber \\
& & \qquad \Tr _E P \Big (  \ket{\Psi_{SEA}(t_n)} \Big ) \cdot  \nonumber \\
& & \left(\left( \mathcal{F}_{\rho_{SA}(t_n)} \cdot \prod _{j=2}^n
 N_{SA^\prime}^{(j)} \mathcal{F}_{\rho_{SA}(t_{j-1})}
\right) \left( N_{SA^\prime}^{(1)} \right) \right)^\dagger \otimes I_{S} \nonumber \\
&=& \left(\left( \mathcal{F}_{\rho_{SA}(t_n)} \cdot \prod _{j=2}^n
 N_{SA^\prime}^{(j)} \mathcal{F}_{\rho_{SA}(t_{j-1})}
\right) \left( N_{SA^\prime}^{(1)} \right) \right) \otimes I_{S} \cdot \nonumber \\
& & \qquad \Tr _E P \Big (  \ket{\tilde{\Psi}_{SEA}(t_n)} \Big ) \cdot \nonumber \\
& & \left(\left( \mathcal{F}_{\rho_{SA}(t_n)} \cdot \prod _{j=2}^n
 N_{SA^\prime}^{(j)} \mathcal{F}_{\rho_{SA}(t_{j-1})}
\right) \left( N_{SA^\prime}^{(1)} \right) \right)^\dagger \otimes I_{S} \nonumber \\
&=&  \Tr _E P \left( \prod_{j=1}^n \left( N_{SA^\prime}^{(j)} \tilde{U}_{SE}^{(j)}\right)
 \ket{\tilde{\Psi}_{SEA}}\right),
\end{eqnarray}
where Eq.~(\ref{eq:psi_j=1_left_n_SA}) of Lemma \ref{ME:lemma4} is used in the first and third
equalities. The above equation is nothing but Eq.~(\ref{A2_eq_definition_of_equivalence_2}) in
Lemma \ref{lemma:sec_problem_setting}, and thus the ``{\it if}" part of the theorem holds. \hfill
$\blacksquare$

\begin{figure}
\includegraphics[scale=0.75]{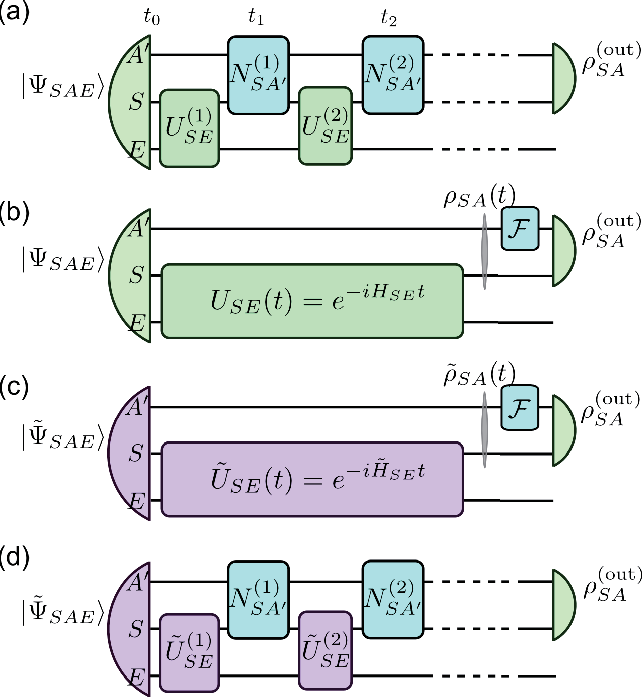}
\caption{(Color online) A diagrammatic demonstration of the proof of Theorem 1. The nontrivial
part of the proof, which is $\rho_{SA}(t)=\tilde \rho_{SA}(t) \Rightarrow (d_E, \ket{\Psi_{SEA}},
H_{SE})\equiv (\tilde{d}_E, \ket{\tilde{\Psi}_{SEA}}, \tilde{H}_{SE})$, is shown. The operations
$N_{SA^\prime}^{(j)}$ on $SA^\prime$ are applied instantaneously at time $t_j$ and
$U_{SE}^{(j)}=\exp[-iH_{SE}(t_j-t_{j-1})]$. Thin grey ovals are inserted in (b) and (c) simply to
emphasise that the state at that instance is $\rho_{SA}(t)$ (or $\tilde \rho_{SA}(t)$). Using
Lemmas \ref{ME:lemma3} and \ref{ME:lemma4}, operations $N_{SA^\prime}^{(j)}$ in (a) can be merged
to an operation $\mathcal{F}$ on the system $A^\prime$ only, as in (b). By assumption, there is
another situation (c) that leads to $\tilde \rho_{SA}(t)$ which is equal to $\rho_{SA}(t)$ in (b).
Since $\mathcal{F}$ can be decomposed to the sequence of $N_{SA^\prime}^{(j)}$ as in (d),
$\rho_{SA}(t)=\tilde \rho_{SA}(t)$ implies that we cannot distinguish the situations (a) and (d),
hence the equivalence.} \label{fig:ProofOfTheorem1}
\end{figure}

\textbf{The flow of the proof of Lemma \ref{ME:lemma4} and Theorem \ref{MEtheorem} is also
depicted in Fig. \ref{fig:ProofOfTheorem1}, which, together with Fig. \ref{fig:ProofOfLemma3},
would help readers understand more easily. In Fig. \ref{fig:ProofOfTheorem1}, the reduced density
operator in Eq. (\ref{ProofMETheorem}) is denoted as $\rho_{SA}^\mathrm{(out)}$.}

Theorem \ref{MEtheorem} leads to the following corollary.
\begin{Corollary}\label{corollary_1}
Suppose $(d_E, \ket{\Psi_{SEA}}, H_{SE})$ satisfies the ME condition in $[t_0, t_\infty)$, and
$d_E < +\infty$. Then, $(d_E, \ket{\Psi_{SEA}}, H_{SE}) \equiv (\tilde{d}_E,
\ket{\tilde{\Psi}_{SEA}}, \tilde{H}_{SE})$ in $[t_0, t_\infty)$, and $\tilde{d}_E < +\infty$ imply
that $(d_E, \ket{\Psi_{SEA}(\tilde{t}_0)}, H_{SE})
 \equiv (\tilde{d}_E, \ket{\tilde{\Psi}_{SEA}(t_0)}, \tilde{H}_{SE})$
in any time interval $[\tilde{t}_0, \tilde{t}_\infty)$.
\end{Corollary}

Roughly speaking, what this claims is as follows. Suppose there are two triples, the one with $E$
and the other with $\tilde{E}$. If the ME condition is fulfilled by one of them and both give rise
to the identical natural time evolution on $SA$ for a finite duration, regardless of its length,
then both triples are equivalent; that is, any active quantum operations on $SA$ cannot reveal the
difference between them.

{\bf (Proof)} \\
Since both $d_E$ and $\tilde{d}_E$ are finite, both $\Tr_E P \left(\ket{\Psi_{SEA}(t)}\right)$ and
$\Tr_E P \left(\ket{\tilde{\Psi}_{SEA}(t)}\right)$ are analytical functions and coincide on $[t_0,
t_\infty)$. Thus, by analytical continuation, $\Tr_E P\left(\ket{\Psi_{SEA}(z)}\right) =\Tr_E P
\left(\ket{\tilde{\Psi}_{SEA}(z)}\right)$ holds for all complex numbers $z$. Together with Theorem
\ref{MEtheorem}, we reach the statement of the corollary.
\hfill $\blacksquare$ \\

\subsection{The ME condition implies maximal entanglement between $A$ and $SE$}\label{sec:pme_subsec:pme_is_mes}
In the previous subsection, we have suggested that when the triple $(d_E, \ket{\Psi_{SEA}},
H_{SE})$ satisfies the ME condition in $[t_0, t_\infty)$, there exists an equivalent triple in
which $\ket{\Phi_{EA_2(t)}}$ can be chosen as a maximally entangled state on the entire space of
$\Hi_E \otimes \Hi_A$. Now we shall prove it as a theorem in this subsection.
\begin{Theorem}\label{Theorem:pme_is_mes}
 Suppose $(d_E, \ket{\Psi_{SEA}}, H_{SE})$ satisfies the
 ME condition in $[t_0, t_\infty)$, and $d_E < + \infty$. Then,
 there exists a triple $(\tilde{d}_E, \ket{\tilde{\Psi}_{SEA}}, \tilde{H}_{SE})$
such that $(\tilde{d}_E, \ket{\tilde{\Psi}_{SEA}}, \tilde{H}_{SE}) \equiv (d_E,
 \ket{\Psi_{SEA}}, H_{SE})$ in $[t_0, t_\infty)$, and $\ket{\tilde{\Psi}_{SEA}}$ is
 a maximally entangled state on the full space of $SEA$ with respect to the partition between
 $SE$ and $A$; that is, $\Tr _A P\left(
 \ket{\tilde{\Psi}_{SEA}}\right)= \frac{1}{d_Sd_E}I_S \otimes I_E$.
\end{Theorem}

\begin{figure}
\includegraphics[scale=0.67]{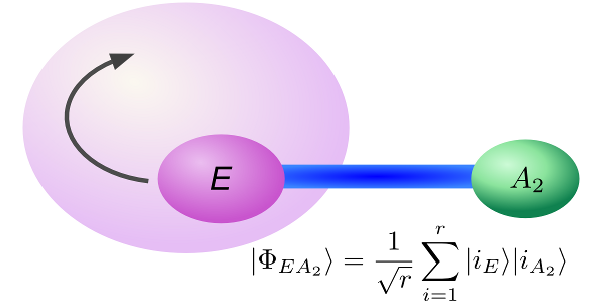}
\caption{(Color online) A situation in which the ME condition may be satisfied, but the support of
the state $\ket{\Phi_{EA_2}}$ in $E$ moves around in a larger space. Theorem
\ref{Theorem:pme_is_mes} tells that there should be an equivalent triple $(\tilde{d}_E,
\ket{\tilde{\Psi}_{SEA}}, \tilde{H}_{SE})$, for which $\mathrm{supp}\rho_E(t)$ matches the whole
$E$ once the ME condition is fulfilled: $E$ can be seen stationary, rather than moving as in this
figure. }\label{fig:theorem2}
\end{figure}

Let us sketch the idea of its proof: For $(d_E, \ket{\Psi_{SEA}}, H_{SE})$ satisfying the ME
condition, $\ket{\Psi_{SEA}}$ is a maximally entangled state on the full space of $S,E,$ and $A$
if and only if $d_E = {\rm rank} \rho_E(t)$, where $\rho_E(t)
:=\Tr_{SA}P\left(e^{-iH_{SE}(t-t_0)}\ket{\Psi_{SEA}}\right)$. If the rank of $\rho_E(t)$ is
smaller than $d_E$, ${\rm supp} \rho_E(t)$ may move around within $E$ as time proceeds (see Fig.
\ref{fig:theorem2} for an intuitive illustration). In such a case, the property of maximally
entangled pairs, Eq.~(\ref{mirroring1}) in Sec.\ref{sec:main_results}, cannot be used for our
Hamiltonian identification purpose, because $H_{SE}$ might change according to the time evolution
of the basis on the $E$ side. Thus, in order to prove the theorem, we need to show the existence
of Hamiltonian $H^\prime_{SE}$ such that $(d_E, \ket{\Psi_{SEA}}, H_{SE}) \equiv (d_E,
\ket{\Psi_{SEA}}, H^\prime_{SE})$ and $\left[ H^\prime_{SE}, \rho_{SE}(t_0)\right]=0$, where
$\rho_{SE}=\Tr_A \ket{\Psi_{SEA}}\bra{\Psi_{SEA}}$. With such a $H^\prime_{SE}$,
$\mathrm{supp}\rho_E(t)$ can be seen stationary.

To this end, we restrict the space $\Hi_E$ to ${\rm supp} \rho'_E(t)$, or equivalently, we define
$\tilde{d}_E$, $\ket{\tilde{\Psi}_{SEA}}$, and $\tilde{H}_{SE}$ as $\tilde{d}_E:= {\rm
rank}\rho_E(t_0)$, $\ket{\tilde{\Psi}_{SEA}}:= \ket{\Psi_{SEA}}$, and
$\tilde{H}_{SE}:=(I_{SA}\otimes P_E(t_0))H'_{SE}(I_{SA}\otimes P_E(t_0))$, respectively, where
$P_E(t):= {\rm rank}\rho^\prime_E(t) \cdot \rho^\prime_E(t)$ is a projector onto the support of
$\rho_E$. Then, the triple $(\tilde{d}_E, \ket{\tilde{\Psi}_{SEA}}, \tilde{H}_{SE})$ satisfies the
desired conditions: $(\tilde{d}_E, \ket{\tilde{\Psi}_{SEA}}, \tilde{H}_{SE}) \equiv (d_E,
\ket{\Psi_{SEA}}, H_{SE})$ in $[t_0, t_\infty)$, and $\ket{\tilde{\Psi}_{SEA}}$ is a maximally
entangled state on the full space of $S\tilde{E}A$.  Therefore, the proof of the theorem can be
reduced to constructing Hamiltonian $H^\prime_{SE}$ so that $(d_E, \ket{\Psi_{SEA}}, H_{SE})
\equiv (d_E, \ket{\Psi_{SEA}}, H^\prime_{SE})$ and $\left[ H^\prime_{SE},
\rho_{SE}(t_0)\right]$=0. In fact, we shall see that the ME condition enables us to do so.

In order to construct $H^\prime_{SE}$, we need more detailed description of $H_{SE}$. In general,
a Hamiltonian $H_{SE}$ can be decomposed as
\begin{equation}\label{eq_HSE_decomposition}
 H_{SE} = I_S \otimes H_{\rm id}+\sum_{\alpha=1}^{d_S^2-1} \sigma_\alpha \otimes H_\alpha,
\end{equation}
where $\{I_S\} \bigcup \{ \sigma_\alpha\}_{\alpha=1}^{d_S^2-1}$ forms an orthogonal basis of a
real space of all Hermitian operators on $\Hi_S$. When $(d_E, \ket{\Psi_{SEA}}, H_{SE})$ satisfies
the ME condition, the state on $SE$ is proportional to $I_S \otimes P_E(t)$, where $P_E(t)$ is a
projector-valued function on $E$. Then, the Schroedinger equation for $\ket{\Psi_{SEA}(t)}$
becomes
\begin{align}
&  i \frac{d}{dt} I_S \otimes P_E(t)\\
 =& \left[ H_{SE}, I_S \otimes P_E(t)
 \right] \nonumber \\
= & I_S \otimes \left[ H_{\rm id}, P_E(t) \right] + \sum_{\alpha=1}^{d_S^2-1}\sigma_\alpha \otimes
\left[ H_\alpha, P_E(t) \right],
\end{align}
for all $t \in \mathbb{R}$.
Thus, $H_{\rm id}$ and $H_{\alpha}$ satisfy
\begin{align}
 i \frac{d}{dt} P_E(t) &= \left[ H_{\rm id}, P_E(t) \right], \label{sec:pme_subsec:pme_is_mes_eq_vonneumann} \\
0 &= \left[ H_\alpha, P_E(t) \right] \quad \left( \forall \alpha \right).
 \label{sec:pme_subsec:pme_is_mes_eq_vonneumann2}
\end{align}
That is, only $H_{\rm id}$ is responsible for the time evolution of $\rho_E(t)=P_E(t)/{\rm
rank}\rho_E(t_0)$. Therefore, a modification to $H_{\rm id}$ may lead to  desired Hamiltonian
$H^\prime_{SE}$. Suppose that the desired modification is described as $H_{\rm id} \rightarrow
H_{\rm id} - H_{\rm id}^\prime$ with a Hermitian operator $H_{\rm id}^\prime$. Such a $H_{\rm
id}^\prime$ is shown to exist and has properties as in the following lemma.
\begin{Lemma}\label{sec:pme_subsec:pme_is_mes lemma}
Suppose $\Hi_E$ is a finite-dimensional Hilbert space, and suppose a projector-valued function
$P_E(t)$ on $\Hi_E$, and Hamiltonians $H_{\rm id}$ and $\left\{ H_\alpha \right\}_{\alpha=1}^c$ on
$\Hi_E$ satisfies Eqs. (\ref{sec:pme_subsec:pme_is_mes_eq_vonneumann}) and
(\ref{sec:pme_subsec:pme_is_mes_eq_vonneumann2}) for all $t \in I_t$, where $I_t$ is an open
subset of $\mathbb{R}$. Then, there exists Hamiltonian $H'_{\rm id}$ that satisfies the following
equations:
\begin{align}
& \left[ H_\alpha, H^\prime_{\rm id} \right]=0 \ \left( \forall \alpha \right) \label{sec:pme_subsec:pme_is_mes_h'01}\\
& \left[ H_{\rm id}, H^\prime_{\rm id} \right]=0, \label{sec:pme_subsec:pme_is_mes_h'02}\\
& \left[ H_{\rm id}, P_E(t) \right]=\left[ H'_{\rm id}, P_E(t) \right], \quad (t \in
 I_t).\label{sec:pme_subsec:pme_is_mes_h'03}
\end{align}
\end{Lemma}
{\bf (Proof)}\\
We generate a sequence of triples $\left\{ \left( \mathfrak{g}_n, G_n, H_{\rm id}^{(n)}
\right)\right\}_{n=1}^\infty$, where $H_{\rm id}^{(n)}$ are Hamiltonians, $\mathfrak{g}_n$ are
linear Lie algebras on $\Hi_E$, and $G_n$ are Lie groups \cite{GOODMAN}, according to the
following procedure. Suppose $H_\alpha$ have spectral decompositions as $H_\alpha= \sum
_{j}h_j^{(\alpha)}P_j^{(\alpha)}$, where $h_j^{(\alpha)} \neq h_{j^\prime}^{(\alpha)}$ for all $j
\neq j^\prime$, and $P_j^{(\alpha)}$ are all projectors. Then, $\mathfrak{g}_1$ is defined as a
Lie algebra whose elements are $\left\{ i P_j^{(\alpha)}\right\}\,\left(\forall j,\alpha \right)$,
$G_1$ is a compact Lie group generated by $\mathfrak{g}_1$, and $H_{\rm id}^{(1)}$ is defined as
\begin{equation}
 H_{\rm id}^{(1)} := \int _{U \in G_1} U H_{\rm id} U^\dagger d\mu \left(U \right ),
\end{equation}
where $\mu(U)$ is a Haar measure \cite{GOODMAN,HAAR} on $G_1$. Starting from $\left
(\mathfrak{g}_1, G_1, H_{\rm id}^{(1)} \right )$, we recurrently define $\left ( \mathfrak{g}_n,
G_n, H_{\rm id}^{(n)} \right )$ as follows: Suppose $H_{\rm id}-H_{\rm id}^{(n)}$ has a spectral
decomposition as $H_{\rm id}-H_{\rm id}^{(n)}= \sum _j h_jQ_j^{(n)}$ , where $Q_j^{(n)}$ are
projectors. Then, we let $\mathfrak{g}_{n+1}$ and $G_{n+1}$ be a linear Lie algebra consisting of
$\mathfrak{g}_n \bigcup \left\{ iQ_j^{(n)}\right\}$ $\left( \forall j \right)$ and a compact Lie
group generated by $\mathfrak{g}_{n+1}$. Similarly as above,  $H_{\rm id}^{(n+1)}$ is defined as
\begin{equation}
 H_{\rm id}^{(n+1)} := \int _{U \in G_{n+1}} U H_{\rm
  id}^{(n)} U^\dagger d\mu
  \left(U \right ).
\end{equation}
Then, we can prove the following equations for all $t \in I_t$ (The time $t$ will denote $t\in
I_t$ hereafter):
\begin{align}
& [H_\alpha, H_{\rm id}^{(n)}] =0, \quad (\forall \alpha, n) \label{sec:pme_subsec:pme_is_mes h_alpha h_o^n}\\
& [H_\mathrm{id}-H_{\rm id}^{(n)}, H_{\rm id}^{(n+1)}] =0, \quad (\forall n) \label{sec:pme_subsec:pme_is_mes h_0-h_0^n}\\
& [H_{\rm id}, P_E(t)]=[H_{\rm id}^{(n)}, P_E(t)], \quad (\forall n). \label{eq:h_0_and_p_e_t}
\end{align}
Note that Eq. (\ref{eq:h_0_and_p_e_t}) can be reexpressed as $[Q_j^{(n)},P_E(t)]=0$ by subtracting
the RHS from the LHS.

Let us first show the following two equations:
\begin{align}
& \left[ P_j^{(\alpha)}, H_{\rm id}^{(n)} \right]=0, \quad (\forall j, \alpha,
 n) \label{sec:pme_subsec:pme_is_mes p_alpha h_o^n}
\end{align}
and
\begin{align}
& \left[ Q_j^{(n)}, H_{\rm id}^{(n+1)} \right]=0 \quad (\forall j, n),
 \label{sec:pme_subsec:pme_is_mes q_j h_o^n}
\end{align}
from which Eqs. (\ref{sec:pme_subsec:pme_is_mes h_alpha h_o^n}) and
(\ref{sec:pme_subsec:pme_is_mes h_0-h_0^n}) are directly obtained. Since $U_\delta := I_E -
P_j^{(\alpha)} + e^{i\delta}P_j^{(\alpha)}$ is in $G_n$ for all $\delta\,(\mathrm{mod}\,2\pi)$,
$H_{\rm id}^{(n)}$ satisfies
\begin{align*}
 \quad&  H_{\rm id}^{(n)} \\
 =& \int _{\delta \in [0,2\pi]} \left( I-P_j^{(\alpha)} +
 e^{i\delta} P_j^{(\alpha)} \right) H_{\rm id}^{(n)}  \\
& \qquad \cdot \left( I-P_j^{(\alpha)} +
 e^{-i\delta} P_j^{(\alpha)} \right) \frac{d\delta}{2\pi} \\
=& (I-P_j^{(\alpha)}) H_{\rm id}^{(n)}(I-P_j^{(\alpha)}) +
 P_j^{(\alpha)}H_{\rm id}^{(n)}P_j^{(\alpha)},
\end{align*}
which implies
\begin{align*}
P_j^{(\alpha)}H_{\rm id}^{(n)} = P_j^{(\alpha)}H_{\rm
 id}^{(n)}P_j^{(\alpha)} = H_{\rm id}^{(n)}P_j^{(\alpha)}.
\end{align*}
Thus, Eq. (\ref{sec:pme_subsec:pme_is_mes p_alpha h_o^n}) holds. Similarly, we can prove Eq.
(\ref{sec:pme_subsec:pme_is_mes q_j h_o^n}) from the fact $U_\delta := I_E - Q_j^{(n)} +
e^{i\delta}Q_j^{(n)}$ in $G_{n+1}$.

We now prove Eq. (\ref{eq:h_0_and_p_e_t}) by induction. First, we consider the case $n=1$.
Equation (\ref{sec:pme_subsec:pme_is_mes_eq_vonneumann2}) implies that $P_j^{(\alpha)}$ satisfies
$\left[ P_j^{(\alpha)}, P_E(t) \right]=0$. Then, by differentiating this equation and using Eq.
(\ref{sec:pme_subsec:pme_is_mes_eq_vonneumann}), we obtain
\begin{align}\label{sec:pme_subsec:pme_is_mes_eq_2012071901}
\left[ P_j^{(\alpha)}, \left [H_{\rm id}, P_E(t) \right ] \right] = 0
\end{align}
 for
all $j$ and $\alpha$. Then, because of the following identities,
\begin{align*}
&\left[ P_j^{(\alpha)}, P_E(t) \right]=0, \\
&\left[P_j^{(\alpha)}, \left [H_{\rm id}, P_E(t) \right ] \right] = 0,\\
& \left[ \left[P_{i_1}^{(\alpha_1)},
P_{i_2}^{(\alpha_2)}\right], P_{i_3}^{(\alpha_3)}\right] + \left[ \left[P_{i_2}^{(\alpha_2)},
P_{i_3}^{(\alpha_3)}\right], P_{i_1}^{(\alpha_1)}\right] \\
& + \left[ \left[P_{i_3}^{(\alpha_3)},
P_{i_2}^{(\alpha_2)}\right], P_{i_1}^{(\alpha_1)}\right] =0,
\end{align*}
the last of which is the Jacobi identify, all operators written in the form of
\begin{equation}\label{sec:pme_subsec:pme_is_mes term 1}
\left  [ \cdots \left [ \left [P_{i_1}^{(\alpha_1)}, P_{i_2}^{(\alpha_2)} \right ], P_{i_3}^{(\alpha_3)} \right
 ]\cdots,  P_{i_k}^{(\alpha_k)} \right ].
\end{equation}
commute with $P_E(t)$ and $\left [ H_{\rm id},  P_E(t) \right ]$. Thus, since an arbitrary $X \in
\mathfrak{g}_1$ can be written as a linear combination of the terms in the form of Eq.
(\ref{sec:pme_subsec:pme_is_mes term 1}), any $X \in \mathfrak{g}_1$ satisfies $\left[ X, P_E(t)
\right]=0$ and $\left[ X, \left [ H_{\rm id},  P_E(t) \right ]\right]=0$.

Since $G_1$ is a compact Lie group, for a given $U \in G_1$, there exists an operator $X \in
\mathfrak{g}_1$ satisfying $U= \exp X$. Thus, an arbitrary $U \in G_1$ satisfies
\begin{align}
 \left[ U,
P_E(t) \right]=0  \label{sec:pme_subsec:pme_is_mes u pet =0}
 \end{align}
and
\begin{align}\label{sec:pme_subsec:pme_is_mes u h0 pet =0}
 \left[ U, \left [ H_{\rm id},  P_E(t) \right ]\right]=0.
\end{align}
Hence, by using Eqs. (\ref{sec:pme_subsec:pme_is_mes u pet =0}) and
(\ref{sec:pme_subsec:pme_is_mes u h0 pet =0}), we can derive the following equation for all $U \in
G_1$:
\begin{align*}
 \left[ H_{\rm id}, P_E(t) \right] &= U  \left[ H_{\rm id}, P_E(t) \right]U^\dagger \\
&= \left[ U H_{\rm id} U^\dagger, P_E(t) \right].
\end{align*}
By integrating the above equation on $G_1$ with a Haar measure $\mu (U)$, we obtain Eq.
(\ref{eq:h_0_and_p_e_t}) for $n=1$, i.e., $\left[ H_{\rm id}, P_E(t) \right] = \left[ H_{\rm
id}^{(1)}, P_E(t) \right]$.

Second, following a discussion similar to the case of $n=1$, we prove Eq. (\ref{eq:h_0_and_p_e_t})
for $n+1$ assuming that Eq. (\ref{eq:h_0_and_p_e_t}) holds for all $k \le n$, i.e., $\left[
Q_j^{(k)}, P_E (t) \right] = 0$ for all $j$ and $k \le n$. As in the case of $P_j^{(\alpha)}$,
this implies
\begin{equation}\label{sec:pme_subsec:pme_is_mes_eq_2012071902}
    \left[ Q_j^{(k)}, \left [H_{\rm id}, P_E (t) \right] \right] = 0
\end{equation}
for all $j$ and $k \le n$. Then, since $\left\{ iQ_j^{(k)} \right \}_{j}$ with all $k \le n$ and
$\left\{ iP_j^{(\alpha)}\right\}_{j, \alpha}$ generate $G_{n+1}$, Eqs.
(\ref{sec:pme_subsec:pme_is_mes_eq_2012071901})  and
(\ref{sec:pme_subsec:pme_is_mes_eq_2012071902}) imply that an arbitrary $X \in \mathfrak{g}_{n+1}$
satisfies $\left[ X, P_E(t) \right]=0$ and $\left[ X, \left [ H_{\rm id},  P_E(t) \right
]\right]=0$. Therefore, as in the case of $n=1$,  we deduce $\left[ H_{\rm id}, P_E(t) \right] =
\left[ H_{\rm id}^{(n+1)}, P_E(t) \right]$.

Equations (\ref{sec:pme_subsec:pme_is_mes h_alpha h_o^n}), (\ref{sec:pme_subsec:pme_is_mes
h_0-h_0^n}), and (\ref{eq:h_0_and_p_e_t}) can be used in the final step to prove the lemma. Since
all $\mathfrak{g}_n$ are Lie subalgebras of $\mathfrak{su} (d_E)$, and thus finite-dimensional,
there exists an $N \in \mathbb{N}$ such that $\mathfrak{g}_n= \mathfrak{g}_{n+1}$ for all $n \ge
N$ (note $\mathfrak{g}_n \subset \mathfrak{g}_{n+1}$). Therefore, we have
\begin{align}
& [H_\alpha, H_{\rm id}^{(N)}] =0, \label{eq:h_alpha_comm_h_id^N}\\
& [H_{\rm id}-H_{\rm id}^{(N)}, H_{\rm id}^{(N)}] =0, \label{eq:h0-h0^n_2}\\
& [H_{\rm id}, P_E(t)]=[H_{\rm id}^{(N)}, P_E(t)]. \label{eq:h_0_comm_p_e_2}
\end{align}
Taking $H'_{\rm id} := H_{\rm id}^{(N)}$ proves the lemma. \hfill $\blacksquare$

Let us proceed to the proof of Theorem \ref{Theorem:pme_is_mes}. \\
{\bf (Proof of Theorem \ref{Theorem:pme_is_mes})} \\
Suppose $(d_E, \ket{\Psi_{SEA}}, H_{SE})$ satisfies the ME condition in $[t_0, t_\infty)$, i.e.,
$P_E(t) := r\Tr _{SA}P\left( \ket{\Psi _{SEA}(t)}\right)$ is a projector-valued function, where $r
:= \dim {\rm supp} \Tr_{SA}P\left( \ket{\Psi_{SEA}}\right)$. Therefore, Eqs.
(\ref{sec:pme_subsec:pme_is_mes_eq_vonneumann}) and
(\ref{sec:pme_subsec:pme_is_mes_eq_vonneumann2}) hold with $H_{\rm id}$ and
 $\left\{ H_\alpha \right\}_{\alpha=1}^c$
defined in Eq. (\ref{eq_HSE_decomposition}). Lemma \ref{sec:pme_subsec:pme_is_mes lemma} then
tells that there exists a Hermitian operator $H'_{\rm id}$ on $\Hi_E$ that satisfies Eqs.
(\ref{sec:pme_subsec:pme_is_mes_h'01})-(\ref{sec:pme_subsec:pme_is_mes_h'03}) for all $I_t \subset
\mathbb{R}$. We now define $H'_{SE}$ as
\begin{equation}\label{eq:def_h'se}
 H'_{SE} := H_{SE}- I_S \otimes H'_{\rm id}.
\end{equation}
Then,  $\left[ H'_{SE}, I_S \otimes H'_{\rm id} \right]$ and $\left[
H'_{SE}, I_S \otimes P_E(t) \right]$  can be computed as
\begin{align}\label{h'se_h'0}
 \left[H'_{SE},  I_S \otimes H'_{\rm id} \right] =&  I_S \otimes [H_{\rm
 id}-H'_{\rm id}, H'_{\rm id}] \nonumber \\
=&0
\end{align}
and
\begin{align}\label{sec:pme_subsec:pme_is_mes h'_se p_e(t)}
 \left[ H'_{SE}, I_S \otimes P_E(t) \right] & = I_S \otimes \left[ H_{\rm
 id}-H'_{\rm id},
 P_E(t)\right] \nonumber \\
&= 0,
\end{align}
where we have used Eqs. (\ref{sec:pme_subsec:pme_is_mes_h'01}) and
(\ref{sec:pme_subsec:pme_is_mes_h'02}) in Eq.(\ref{h'se_h'0}), and Eq.
(\ref{sec:pme_subsec:pme_is_mes_h'03}) in Eq. (\ref{sec:pme_subsec:pme_is_mes h'_se p_e(t)}).
Then, by defining $\ket{\Psi'_{SEA}(t)} := \exp \left(-i H'_{SE}(t-t_0) \right)\ket{\Psi_{SEA}}$,
we obtain
\begin{align}
& \Tr _E P\left( \ket{\Psi_{SEA}(t)}\right) \nonumber \\
=& \Tr _E \exp \left( -iH_{SE}(t-t_0) \right)P\left(
 \ket{\Psi_{SEA}}\right) \exp \left(  iH_{SE}(t-t_0) \right)\nonumber \\
=& \Tr _E \exp \left( -i\left(I_S \otimes H'_{\rm id} \right)(t-t_0) \right) \exp
 \left( -iH'_{SE}(t-t_0) \right) \nonumber \\
& \quad \cdot P\left( \ket{\Psi_{SEA}}\right) \exp \left(
 iH'_{SE}(t-t_0) \right) \nonumber \\
& \quad \cdot \exp \left( i\left(I_S \otimes H'_{\rm id} \right)(t-t_0) \right) \nonumber \\
=& \Tr _E \exp \left( -iH'_{SE}(t-t_0) \right)P\left(
 \ket{\Psi_{SEA}}\right) \exp \left(  iH'_{SE}(t-t_0) \right)\nonumber \\
=& \Tr _E P\left( \ket{\Psi'_{SEA}(t)}\right),
\end{align}
where we have used Eqs. (\ref{eq:def_h'se}) and (\ref{h'se_h'0}) in the second and third
equalities, respectively. Also, since the state on $SE$ can be written as
\begin{align}
& Tr _A P\left( \ket{\Psi^\prime_{SEA}(t)}\right) \nonumber \\
=& \frac{1}{rd_S} \exp \left( -i H^\prime_{SE}(t-t_0)\right) \nonumber \\
& \quad \cdot \left( I_S \otimes P_E(t_0)
 \right) \exp\left( i H^\prime_{SE}(t-t_0)\right) \nonumber \\
=& \frac{1}{rd_S} I_S \otimes P_E(t_0),
\end{align}
the state $\Tr _{SA} P\left( \ket{\Psi'_{SEA}(t)}\right) = \frac{1}{r}P_E(t_0)$ is stationary,
thus, does not move around in a larger space. Therefore, by defining a Hilbert space
$\tilde{\Hi}_E$, a state $\ket{\tilde{\Psi}_{SEA}}$, and a Hamiltonian $\tilde{H}_{SE}$ on $\Hi_S
\otimes \tilde{\Hi}_E$ as
\begin{align}
 \tilde{\Hi}_E & := {\rm supp} P_E(t_0), \\
\ket{\tilde{\Psi}_{SEA}} &:= \ket{\Psi^\prime_{SEA}
 (t_0)}=\ket{\Psi_{SEA}}, \\
\tilde{H}_{SE} &:= I_S \otimes P_E(t_0)H^\prime_{SE}I_S \otimes P_E(t_0),
\end{align}
the triple $\left(\tilde{d}_E, \ket{\tilde{\Psi}_{SEA}}, \tilde{H}_{SE} \right)$ satisfies the
statement of the theorem, where $\tilde{d}_E := \dim \tilde{\Hi}_E = {\rm rank} P_E(t_0)$. \hfill
$\blacksquare$

An important corollary of Theorem \ref{Theorem:pme_is_mes} follows:
\begin{Corollary} \label{sec:pme_subsec:pme_is_mes_corollary}
 Suppose $(d_E, \ket{\Psi_{SEA}}, H_{SE})$ satisfies the
 ME condition in $[t_0, t_\infty)$, and $\dim d_E < + \infty$.
Then, $(d_E, \ket{\Psi_{SEA}(t_0')}, H_{SE})$ with $\ket{\Psi_{SEA}(t'_0)}$ being given by Eq.
(\ref{eq:def_psi_t}) also satisfies the
 ME condition in $[t_0', t_\infty')$ for all $t_0'$ and $t_\infty'$
 $\left( t_0' < t_\infty' \right)$.
\end{Corollary}
{\bf (Proof)} \\
\quad  Theorem \ref{Theorem:pme_is_mes} guarantees that there exists $(\tilde{d}_E,
\ket{\tilde{\Psi}_{SEA}}, \tilde{H}_{SE})$ such that $(d_E, \ket{\Psi_{SEA}}, H_{SE}) \equiv
(\tilde{d}_E, \ket{\tilde{\Psi}_{SEA}}, \tilde{H}_{SE})$ in $[t_0, t_\infty)$ and
$\ket{\tilde{\Psi}_{SEA}}$ is a MES with respect to the partition $SE|A$. Then,  for all $t_0'$
and $t_\infty'$ satisfying $t_0' < t_\infty'$, $(d_E, \ket{\Psi_{SEA}(t'_0)}, H_{SE}) \equiv
(\tilde{d}_E, \ket{\tilde{\Psi}_{SEA}(t'_0)}, \tilde{H}_{SE})$ in $[t'_0, t'_\infty)$, as stated
in Corollary \ref{corollary_1}. On the other hand, since $\ket{\tilde{\Psi}_{SEA}}$ is a MES,
$(\tilde{d}_E, \ket{\tilde{\Psi}_{SEA}(t'_0)}, \tilde{H}_{SE})$ satisfies the ME condition in any
time intervals $[t'_0, t'_\infty)$  $\left( t'_0 < t'_\infty \right)$. Recalling the definition of
the triple equivalence (and also the ``if" part of Theorem \ref{MEtheorem}), this leads to the
statement of the corollary.
\hfill $\blacksquare$ \\
\quad This corollary tells that if our system satisfies the ME condition for a non-zero time
period, no matter how short it is, it will always satisfy the ME condition from then on.

\section{Tomography of the environment under the ME condition}\label{sec:tomography}
In this section, we show that, when the entire system $SEA$ satisfies the ME condition, it is
possible to specify the equivalence class by performing tomography on joint system $SA$. More
precisely, when our triple $\left(d_E,  \ket{\Psi_{SEA}}, H_{SE} \right)$ satisfies the ME
condition in the time period $[t_0, t_\infty)$, we can reconstruct $\tilde{d}_E$,
$\ket{\tilde{\Psi}_{SEA}}$ and $\tilde{H}_{SE}$ that satisfy $\left(d_E,  \ket{\Psi_{SEA}},
H_{SE}\right) \equiv \left(\tilde{d}_E,  \ket{\tilde{\Psi}_{SEA}}, \tilde{H}_{SE}\right)$ in
$[t_0, t_\infty)$ by simply performing tomography of $SA$ while $SEA$ evolves naturally. As we have explained in Sec. \ref{subsec:control-of-E}, this information is
sufficient to control system $E$ and exploit it as a resource for quantum computation.

Let us start with another corollary of Theorem \ref{Theorem:pme_is_mes}:
\begin{Corollary}\label{sec:pme_subsec:tomography_corollary_1}
Suppose a triple $\left(d_E, \ket{\Psi_{SEA}}, H_{SE} \right)$ satisfies the ME condition in
$[t_0, t_\infty)$. Then, there exist Hilbert spaces $\Hi_{A_1}$ and $\Hi_{A_2}$ and a triple
$\left(\tilde{d}_E, \ket{\tilde{\Psi}_{SEA}}, \tilde{H}_{SE}\right)$, such that $\Hi_A =
\Hi_{A_1}\otimes \Hi_{A_2}$, and $\ket{\tilde{\Psi}_{SEA}}$ can be written as
\begin{equation}\label{sec:pme_subsec:tomography_corollary_1_eq}
 \ket{\tilde{\Psi}_{SEA}}= \ket{\Upsilon_{SA_1}}\otimes \ket{\Upsilon_{EA_2}},
\end{equation}
where $\ket{\Upsilon_{SA_1}} \in \Hi_S\otimes\Hi_{A_1}$ and $\ket{\Upsilon_{EA_2}} \in
\Hi_E\otimes\Hi_{A_2}$ are maximally entangled states, and $\tilde{d}_E = \dim {\rm supp}
\Tr_{A_2} P \left( \ket{\Upsilon_{EA_2}}\right)$. Moreover, we can choose a basis set freely for
both $A_2$ and $E$ such that $\ket{\Upsilon_{EA_2}}$ can be expressed as
\begin{equation}\label{std_max_ent}
 \ket{\Upsilon_{EA_2}}= \frac{1}{\sqrt{\tilde{d}_E}}\sum_{i=1}^{\tilde{d}_E} \ket{i_E}\ket{i_{A_2}}.
\end{equation}
\end{Corollary}
The last part of the corollary states that there is a freedom in the choice of basis for $A_2$ and
$E$ due to the equivalence of observable dynamics induced by Hamiltonians that may be equivalent
up to a local unitary $U_E$.

{\bf (Proof)} \\
The first part of the statement is simply a rephrase of Theorem \ref{Theorem:pme_is_mes}. Thus, we
only prove the second part, which says that $\ket{\Upsilon_{EA_2}}$ can always be written in the
form of Eq. (\ref{std_max_ent}) even if we choose an arbitrary basis. Suppose that $\left(d_E,
\ket{\Psi_{SEA}}, H_{SE}\right)$ satisfies the ME condition in $[t_0, t_\infty)$ and
$\ket{\tilde{\Psi}_{SEA}}$ and $\tilde{H}_{SE}$ are given as $\ket{\tilde{\Psi}_{SEA}} := (I
\otimes U_E) \ket{\Psi_{SEA}}$ and $\tilde{H}_{SE}:= (I_{S}\otimes U_E) H_{SE} (I_S \otimes
U_E^\dagger)$, respectively, where $U_E$ is an arbitrary unitary operator on $\Hi_E$. Then,
$\ket{\Psi_{SEA}(t)} := e^{-iH_{SE}t}\ket{\Psi_{SEA}}$ and $\ket{\tilde{\Psi}_{SEA}(t)}
:=e^{-i\tilde{H}_{SE}t}\ket{\tilde{\Psi}_{SEA}}$ satisfy $\Tr_E P\left(
\ket{\Psi_{SEA}(t)}\right)=\Tr_E\left(\ket{\tilde{\Psi}_{SEA}(t)}\right)$, for all $t\in[t_0,
t_\infty)$, i.e., $\left(d_E, \ket{\Psi_{SEA}}, H_{SE} \right) \equiv \left(\tilde{d}_E,
\ket{\tilde{\Psi}_{SEA}}, \tilde{H}_{SE} \right)$. Hence, the state $\ket{\Psi_{SEA}}$ has a
unitary freedom on $E$, which means that it is possible to choose a basis to express it as Eq.
(\ref{std_max_ent}).
\hfill $\blacksquare$ \\

Therefore, when our system satisfies the ME condition, we can determine the decomposition
$\Hi_A=\Hi_{A_1}\otimes \Hi_{A_2}$, the state $\ket{\Upsilon_{SA_1}}$, and the dimension
$\tilde{d}_E$ satisfying $\tilde{d}_E=\dim {\rm supp} \Tr_{SEA_1} P\left( \ket{\Psi_{SEA}}\right)$
by performing (joint) state tomography on $SA$. Moreover, by redefining $\Hi_A := {\rm supp}
\Tr_{SE} P\left( \ket{\Psi_{SEA}}\right)$, we can assume $\dim \Hi_{A_2} = \dim \Hi_E =
\tilde{d}_E$ and $\dim \Hi_A = d_S \tilde{d}_E$. The above corollary
also implies that we can
always assume system $EA_2$ is in the $\tilde{d}_E\times\tilde{d}_E$-dimensional standard
maximally entangled state (of the form of Eq. (\ref{std_max_ent})).

The remaining task is now to determine the interaction Hamiltonian $H_{SE}$ (finally!). Theorems
\ref{MEtheorem} and \ref{Theorem:pme_is_mes} allow us to state the following:
\begin{Corollary}\label{sec:pme_subsec:tomography_corollary_2}
Suppose a triple $\left(d_E, \ket{\Psi_{SEA}}, H_{SE} \right)$ satisfies the ME condition in
$[t_0, t_\infty)$. If a Hermitian matrix $\tilde{H}_{SE}$ on $\Hi_S \otimes \Hi_E$ satisfies
\begin{align}\label{sec:pme_subsec:tomography_corollary_2_eq}
i \frac{d}{dt} \rho_{SA}(t) = \left[ I_S \otimes \tilde{H}_{SE}^{T}, \rho_{SA}(t) \right]
\end{align}
for all $t$ in a neighbourhood of $t_0$, then $\left(d_E, \ket{\Psi_{SEA}}, H_{SE} \right)\equiv
\left(d_E, \ket{\Psi_{SEA}}, \tilde{H}_{SE} \right)$ in $[t_0, t_\infty)$. In Eq.
(\ref{sec:pme_subsec:tomography_corollary_2_eq}), $\rho_{SA}(t) :=\Tr_E P\left(e^{-iH_{SE}(t-t_0)}
\ket{\Psi_{SEA}}\right)$, the transposition $T$ is taken with respect to the Schmidt basis of
$\ket{\Psi_{SEA}}$ and $\tilde{H}_{SE}^{T}$ is an operator on $\Hi_A$.
\end{Corollary}
Thanks to Corollary 3, we can take it for granted that $\ket{\Psi_{SEA}}$ has the form of Eq.
(\ref{sec:pme_subsec:tomography_corollary_1_eq}). Corollary
\ref{sec:pme_subsec:tomography_corollary_2} implies that, for a given $\rho_{SA}(t)$, which can be
specified by state tomography, all Hermitian matrices $\tilde{H}_{SE}$ satisfying Eq.
(\ref{sec:pme_subsec:tomography_corollary_2_eq}) can be adopted as the interaction Hamiltonian
between systems $S$ and $E$.

Let us now describe how to find such a matrix $\tilde{H}_{SE}$ from the observed data of
$\rho_{SA}(t)$. Since the time evolution of $\rho_{SA}(t)$ is induced by the (finite-dimensional)
matrix $\tilde{H}_{SE}$, there exist a set of real numbers $\left\{ \theta_\alpha
\right\}_{\alpha=1}^L$ and a set of linear operators $\left\{ \rho_\alpha\right\}_{\alpha=0}^L$ on
$\Hi_S \otimes \Hi_A$ such that $\rho_{SA}(t)$ can be written as
\begin{equation}\label{sec:pme_subsec:tomography_eq_rhoSA_t}
 \rho_{SA}(t)=\rho_0 + \sum _{\alpha=1}^L \left( e^{i
                       \theta_\alpha(t-t_0)}\rho_\alpha +
                       e^{-i\theta_\alpha(t-t_0)} \rho_\alpha^\dagger \right),
\end{equation}
where $L$ is at most $d_S \tilde{d}_{E} \left(d_S \tilde{d}_{E}-1 \right)/2$ and  $\rho_0$ is
Hermitian. Setting $t=t_0$, we have
\begin{equation}
 \rho_{SA}(t_0)=\rho_0 + \sum _{\alpha=1}^L \left( \rho_\alpha + \rho_\alpha^\dagger \right).
\end{equation}
Further, differentiating Eq. (\ref{sec:pme_subsec:tomography_eq_rhoSA_t}) $n$ times leads to
\begin{equation}\label{sec:pme_subsection_tomography_eq_dn_dt_n_rhoSA}
\frac{d^n }{dt^n}\rho_{SA}(t)\Big |_{t=t_0} = \sum_{\alpha=1}^L \left\{ \left(i\theta _\alpha
\right)^n \rho_\alpha + \left(-i\theta _\alpha \right)^n \rho_{\alpha}^\dagger \right\}.
\end{equation}
Therefore, we can determine $\left\{ \theta_\alpha \right\}_{\alpha=1}^L$ and $\left\{
\rho_\alpha\right\}_{\alpha=0}^L$ from at most $d_A^2$-th order derivative of $\rho_{SA}(t)$ at
$t=t_0$, which can be obtained experimentally in principle.

The information on $\left\{ \theta_\alpha \right\}_{\alpha=1}^L$ and $\left\{
\rho_\alpha\right\}_{\alpha=0}^L$ allows us to reconstruct $\tilde{H}_{SE}$ so that it satisfies
Eq. (\ref{sec:pme_subsec:tomography_corollary_2_eq}). Here is a lemma that shows a property the
desired matrix should have:
\begin{Lemma}\label{sec:pme_subsec:tomography_lemma_1}
A set of real numbers $\left\{ \theta_\alpha \right\}_{\alpha=1}^L$, a set of linear operators
$\left\{ \rho_\alpha\right\}_{\alpha=0}^L \subset \B \left( \Hi_S \otimes \Hi_A \right)$, and a
Hermitian matrix $H \in \B \left( \Hi_A \right)$ satisfy the following equations:
\begin{align}\label{sec:pme_subsec:tomography_lemma_1_eq1}
[H, \rho_0]&=0 \nonumber \\
[H, \rho_\alpha] &= -\theta_\alpha \rho_\alpha, \quad (1 \le \forall
\alpha \le L,)\nonumber \\
\end{align}
if and only if $\rho_{SA}(t) \in \B\left(\Hi_S\otimes \Hi_A\right)$ given by Eq.
(\ref{sec:pme_subsec:tomography_eq_rhoSA_t}) satisfies
\begin{equation}\label{sec:pme_subsec:tomography_lemma_1 eq2}
 i\frac{d}{dt}\rho_{SA}(t) = \left[ H, \rho_{SA}(t) \right].
\end{equation}
\end{Lemma}
In Eqs. (\ref{sec:pme_subsec:tomography_lemma_1_eq1}) and (\ref{sec:pme_subsec:tomography_lemma_1
eq2}), $I_S$ is omitted for simplicity; that is, $H$ in these equations means $H\otimes I_S$.

{\bf (Proof)} \\
Simply taking the time derivative of Eq. (\ref{sec:pme_subsec:tomography_eq_rhoSA_t}) and then
using Eq. (\ref{sec:pme_subsec:tomography_lemma_1_eq1}) lead to Eq.
(\ref{sec:pme_subsec:tomography_lemma_1 eq2}) to prove the \textit{only if} part. The \textit{if}
part can be shown by substituting Eq. (\ref{sec:pme_subsec:tomography_eq_rhoSA_t}) into Eq.
(\ref{sec:pme_subsec:tomography_lemma_1 eq2}):
\begin{align}\label{for_lemma6}
&  [H,\rho_0]+ \sum _{\alpha=1}^L \Big \{ \left( [H,\rho_\alpha]-\theta_\alpha \rho_\alpha
 \right)e^{i\theta_\alpha t}  \nonumber \\
& \quad + \left( [H,\rho_\alpha^\dagger]+\theta_\alpha\rho_\alpha^\dagger
 \right)e^{-i\theta_\alpha t}\Big \}=0.
\end{align}
Equation (\ref{sec:pme_subsec:tomography_lemma_1_eq1}) follows if we apply $\lim _{T \rightarrow
+\infty}\frac{1}{T}\int_{0}^Tdt e^{i\theta_{\alpha^\prime}t}$ to Eq. (\ref{for_lemma6}).
\hfill $\blacksquare$ \\

Lemma \ref{sec:pme_subsec:tomography_lemma_1} assures that any Hermitian matrix $H$ satisfying Eq.
(\ref{sec:pme_subsec:tomography_lemma_1_eq1}) can be identified as an $\tilde{H}_{SE}$ (taking its
transpose $H^T$). The existence of such a matrix $H$, i.e., a Hermitian matrix that satisfies Eq.
(\ref{sec:pme_subsec:tomography_lemma_1_eq1}), is guaranteed by Corollary
\ref{sec:pme_subsec:tomography_corollary_2}. Therefore, our task now is to find a specific form of
$H$, given the information on $\rho_{SA}(t)$.

To this end, we fix an orthonormal basis of all Hermitian operators $\left\{ A_j
\right\}_{j=1}^{d_S^4\tilde{d}^{2}_E} \subset \B \left( \Hi_S \otimes \Hi_A \right)$, each of
which has the form $A_j=B_j \otimes I_S$, where $\left\{ B_j \right\}_{j=1}^{d^2_S
\tilde{d}^{2}_E} \subset \B \left( \Hi_A \right)$ (Recall that our target Hamiltonian has the form
$H_A\otimes I_S$ and $\rho_{SA_1}$ is maximally entangled.). We also let $\left\{ \epsilon_{jkl}
\right\}_{jkl} \,(j,k,l\in\{1,2,..., d^4_S \tilde{d}_E^{2}\})$ denote a set of structure constants
of the Lie algebra generated by $\left\{ i A_j \right\}_{j=1}^{d^4_S \tilde{d}_E^{2}}$, i.e.,
\begin{equation}
 \left[iA_j, iA_k \right] = \sum _{l=1}^{d^4_S \tilde{d}_E^{2}}\epsilon_{jkl}iA_l.
\end{equation}
Since the basis set $\left\{ A_j \right\}_{j=1}^{d^4_S \tilde{d}^{2}_E}$ is a basis of all
Hermitian operators on $\Hi_S \otimes \Hi_A$, we can uniquely expand the following operators in
terms of $\{A_j\}$:
\begin{align}
\rho_0 &= \sum_{j=1}^{d^4_S \tilde{d}^{2}_E} u_j A_j, \nonumber \\
\rho_\alpha + \rho_\alpha^\dagger &= \sum_{j=1}^{d^4_S \tilde{d}^{2}_E} v_j^{(\alpha)} A_j, \nonumber \\
i \left( \rho_\alpha - \rho_\alpha^\dagger \right) &= \sum_{j=1}^{d^4_S \tilde{d}^{2}_E} w_j^{(\alpha)} A_j, \nonumber \\
\end{align}
where $\left\{ u_j\right\}_{j=1}^{d^4_S \tilde{d}^{2}_E}$, $\left\{ v_j^{(\alpha)}
\right\}_{j=1}^{d^4_S \tilde{d}^{2}_E}$, and $\left\{ w_j^{(\alpha)} \right\}_{j=1}^{d^4_S
\tilde{d}^{2}_E}$ are all real constants for all $\alpha$. Similarly, an arbitrary Hermitian
operator $H$ on $\Hi_A$ can be written as
\begin{equation}
H \otimes I_S = \sum_{j=1}^{d^2_S \tilde{d}^{2}_E} h_j A_j,
\end{equation}
with real constants $\left\{ h_j \right\}_{j=1}^{d^2_S \tilde{d}^{2}_E}$. Then, a necessary and
sufficient condition for $H$, $\left\{ \theta_\alpha \right\}_{\alpha=1}^L$, and $\left \{
\rho_\alpha \right \}_{\alpha=0}^L$ to satisfy Eq. (\ref{sec:pme_subsec:tomography_lemma_1_eq1})
is that $\left\{ h_j \right\}_{j=1}^{d^2_S \tilde{d}^{2}_E}$ satisfies the system of linear
equations with $1 \le l \le d^4_S  \tilde{d}^{2}_E$ and $1 \le \alpha \le L$:
\begin{align}\label{system_of_equations_for_hj}
\sum _{j=1}^{d^2_S \tilde{d}^{2}_E} \left( \sum _{k=1}^{d^4_S \tilde{d}^{2}_E}  \epsilon _{jkl}
 u_k \right) h_j &=0, \nonumber \\
\sum _{j=1}^{d^2_S \tilde{d}^{2}_E} \left( \sum _{k=1}^{d^4_S \tilde{d}^{2}_E}  \epsilon _{jkl}
 v_k^{(\alpha)} \right) h_j &=\theta_\alpha w_k^{(\alpha)}, \nonumber \\
\sum _{j=1}^{d^2_S \tilde{d}^{2}_E} \left( \sum _{k=1}^{d^4_S \tilde{d}^{2}_E}  \epsilon _{jkl}
 w_k^{(\alpha)} \right) h_j &=-\theta_\alpha v_k^{(\alpha)}. \nonumber \\
\end{align}
The system of linear equations (\ref{system_of_equations_for_hj}) has at least one solution, as
for experimentally obtained $\left\{ \theta_\alpha \right\}_{\alpha=1}^L$ and $\left\{ \rho_\alpha
\right\}_{\alpha=0}^L$ there must exist $H$ that satisfies Eq.
(\ref{sec:pme_subsec:tomography_lemma_1_eq1}).
 Eqs. (\ref{system_of_equations_for_hj}) may have
multiple independent solutions. As we have mentioned above, for
all such Hermitian operators $H$, its transpose $H^T$ can be a legitimate Hamiltonian
$\tilde{H}_{SE}$ describing the dynamics of our triple, $\left(\tilde{d}_E,
\ket{\Upsilon_{SA_1}}\otimes \ket{\Upsilon_{EA_2}}, \tilde{H}_{SE} \right)$. Hence, the mission
completed. :-)

\section{A protocol for state-steering toward the fulfillment of the ME  condition}\label{sec:state_steering}
As we have seen in Sec. \ref{sec:pme}, as long as our access is limited to the principal and
ancillary systems $S$ and $A$, the best we can do (from quantum control perspective) is to
determine an equivalence class of triples in the form of $\left(d_E, \ket{\Psi_{SEA}}, H_{SE}
\right)$. The information on $H_{SE}$ thereby obtained is sufficient to exploit environment $E$ as
a resource for quantum computing by actively controlling it via system $S$. In order for this
Hamiltonian identification to work out, the state $\ket{\Psi_{SEA}}$ has to satisfy the ME
condition in our scenario.

Therefore, we need a method to steer any given state on $\mathcal{H}_S\otimes\mathcal{H}_E$
toward such a state that fulfills the ME condition. In this section, we present a protocol for
this task allowing us to  append an extra (ancillary) system $\mathcal{H}_A$.

As we have mentioned in Sec \ref{sec:problem_setting}, it can be taken for granted that the whole
state on $\Hi_S\otimes \Hi_E$ is pure at the beginning. In addition, this initial state on
$\Hi_S\otimes \Hi_E$ is assumed to be the same (fixed), but perhaps an unknown, state after
equilibration. We set $t=0$ when each iteration of the protocol starts. The protocol proceeds by
iterating a block of steps that consist of four major elements, namely the SWAP operation
\cite{NIELSEN} between $S$ and (a subsystem of) $A$, the time evolution driven by the Hamiltonian
$H_{SE}$, the local filtering operation $\mathcal{F}_\mathrm{LF}^A$ on $A$, and state tomography
on $S$ and $A$. The number of iterations of the block is indexed by $C$.

Since we need to perform state tomography on $SA$ at sufficiently high frequency \textbf{to have
the information on the time evolution of $\rho_{SA}$ and quantities related to it, such as $\Delta
E_{SA}$,} and there are non-deterministic operations $\mathcal{F}_\mathrm{LF}^A$, the protocol
needs to be iterated (ideally infinitely) many times by resetting the entire state and the clock.

A quantity $\Delta E_{SA}$, which plays a central role in designing the protocol, is defined as a
functional of state on $\Hi_S\otimes \Hi_A$ as
\begin{equation}\label{Delta_E_SA}
\Delta E_{SA}:=S(\rho_{SA})-S(\rho_A)+\log d_S,
\end{equation}
where $S(\rho)=-\Tr \rho\ln\rho$ is the von Neumann entropy of state $\rho$
\cite{VONNEUMANN,WEHRL1978,OHYA}.

The local filtering operation $\mathcal{F}_\mathrm{LF}^A$ is given as
$\mathcal{F}_\mathrm{LF}^A(\rho)= F_\mathrm{LF}\rho F_\mathrm{LF}$, where $F_\mathrm{LF} :=
\sqrt{\lambda_\textrm{min}\cdot \rho_A^{-1}}$, where $\rho_A^{-1}$ and $\lambda_\textrm{min}$ are
the inverse of $\rho_A$ on its support and the smallest eigenvalue of $\rho_A$. The local
filtering operation succeeds with probability $\lambda_\textrm{min}\cdot \textrm{rank}\rho_A$.
When the local filtering fails, we abort the present protocol and restart it from the beginning.

The protocol proceeds as follows:
\begin{description}
 \item[Step 0.] At $t=0$, i.e., before any iteration of the following steps, there is no ancillary
        system $A$, thus $\dim \Hi_A=1$. Set the counter $C=0$.
 \item[Step 1.] Prepare a standard MES on a pair of new $d_S$-dimensional ancillary systems, $\Hi_{a_1} \otimes \Hi_{a_2}$. Apply
       $SWAP_{Sa_1}$ on $\Hi_S \otimes \Hi_{a_1}$ and relabel the resulting group of $Aa_1a_2$ as a new $A$ system.
 \item [Step 2.] Apply the local filtering operation
        $\mathcal{F}_\mathrm{LF}^A$ on $\Hi_A$ and increase $C$ by one and call this time $t_C$.
        If it fails, carry out the whole protocol from the beginning, initialising the state
        $\ket{\Psi_{SEA}}$ (to an unknown, but the same, state).
 \item [Step 3.] Evaluate $\Delta E_{SA}$ in Eq. (\ref{Delta_E_SA}) through state tomography on
 $SA$ and define $\epsilon_C$ as
\begin{equation}\label{epsilon_C}
\epsilon_C := \frac{1}{2} \textrm{sup}_{} \left\{  \Delta E_{SA}\left(t
        \right) \ | \ t \in \left[t_C, t_C + \Delta t
        \right] \right\}
\end{equation}
during $\left[ t_C, t_C +\Delta t_C \right]$.
 \item[Step 4.] Terminate the protocol if $\epsilon_C=0$; otherwise, let the $SE$ system evolve until
 $\Delta E_{SA}$ becomes larger than (or equal to) $\epsilon_C$ and go back to Step 1.
\end{description}

As mentioned earlier, because state tomography is involved in Step 3, the evaluation of $\Delta
E_{SA}$ can be achieved by iterating all the preceding steps many times. Having obtained $\Delta
E_{SA}$ as a function of time over $[t_C,t_C+\Delta t]$, we set $\epsilon_C$ to be a threshold
that can be attained within this time period. The factor of 1/2 in Eq. (\ref{epsilon_C}) is chosen
merely for convenience to define an achievable threshold.

How long can $\Delta t$ be? It is the time length, within which we obtain $\Delta E_{SA}$ as a
function of time by repeating state tomography. As long as our active controls, such as SWAP and
$\mathcal{F}_\mathrm{LF}^A$, can be performed in an error-free manner,
as we assume in this study, $\Delta t$
can be arbitrary. In order to minimise the overall time length, the
shorter the $\Delta t$ is the
better; however, if $\Delta t$ is too short the local filtering would succeed with only a very
small probability. Therefore, a more realistic length of $\Delta t$ would be the period of
'oscillation' of $\Delta E_{SA}$, which means that the value of $\Delta t$ may vary from time to
time. Nevertheless, this strategy should work, since $\Delta t$ only needs to
have an
appropriate $\epsilon_C$, and its precise, or best, value does not have to be known a priori.

An important observation follows as a theorem:
\begin{Theorem}\label{sec:protocol_theorem_1}
If $d_E:=\dim \Hi_E$ is finite, the protocol halts when $C=d_E$ at latest, i.e., before
$t=d_E\Delta t$. In other words, there exists a natural number $K \le d_E$ such that $\Delta
E_{SA} (t)=0$ for all $t \in \left[ t_K, t_K + \Delta t \right]$, or equivalently, $\epsilon_K=0$.
\end{Theorem}
Let us first see that $\Delta E_{SA}$ at time $t$ is equal to the amount of the increment of
entanglement between $\Hi_A$ and $\Hi_S \otimes \Hi_E$, which is induced when Step 1 of the
protocol is performed at $t$.
\begin{Lemma}\label{lemma:deltaE=Eout-Ein}
Suppose that the joint system $SEA$ at time $t$ in the protocol is described by a state
$\ket{\Psi_{SEA}}$, and $\ket{\Psi_{SEA}^{in}}$ and $\ket{\Psi_{SEA}^{out}}$ are states on $\Hi_S
\otimes \Hi_E \otimes \Hi_A \otimes \Hi_{a_1}\otimes \Hi_{a_2} $ given as
\begin{eqnarray}
 \ket{\Psi_{SEA}^{in}} &:=& \ket{\Psi_{SEA}}\otimes\ket{\Upsilon_{a_1a_2}}, \nonumber \\
 \ket{\Psi_{SEA}^{out}} &:=& SWAP_{Sa_1}\ket{\Psi_{SEA}^{in}}  \nonumber \\
&=& \ket{\Psi_{a_1EA}}\otimes \ket{\Upsilon_{Sa_2}},
\end{eqnarray}
where $\ket{\Upsilon_{a_1 a_2}}$ and $\ket{\Upsilon_{Sa_2}}$ are written in the standard form of
$d_S$-dimensional maximally entangled states on $\Hi_{a_1}\otimes
 \Hi_{a_2}$ and $\Hi_S\otimes \Hi_{a_2}$, respectively.
Then, for $\rho_{SA} := \Tr _E P\left(\ket{\Psi_{SEA}}\right)$ and $\Delta E_{SA}$ defined in Eq.
(\ref{Delta_E_SA}), we have
\begin{align}\label{eq:sec_protocol_lemma_1_eq}
\Delta E_{SA}=E(\ket{\Psi_{SEA}^{out}})-E(\ket{\Psi_{SEA}^{in}},
\end{align}
where $E(\ket{\Psi})$ is the amount of entanglement with respect to the partition between $\Hi_S
\otimes \Hi_E$ and $\Hi_A\otimes \Hi_{a_1}\otimes
 \Hi_{a_2}$ \cite{NIELSEN,HORODECKI2009,BENNETT1996,PLENIO2007}.
\end{Lemma}

({\bf Proof})\\
\begin{align*}
& E(\ket{\Psi_{SEA}^{out}})-E(\ket{\Psi_{SEA}^{in}}) \\
=& E(\ket{\Psi_{Aa_1E}} \otimes \ket{\Upsilon_{Sa_2}}) - E(\ket{\Psi_{SEA}} \otimes \ket{\Upsilon_{a_1a_2}}) \\
=& E(\ket{\Psi_{Aa_1E}} \otimes \ket{\Upsilon_{Sa_2}}) - E(\ket{\Psi_{SEA}}) \\
=& S(\rho_{Aa_1}\otimes \rho_{mix})-S(\rho_{A})\\
=&S(\rho_{SA})-S(\rho_{A})+\log d_S \\
=& \Delta E_{SA}.
\end{align*}
In the RHS of the fourth line, $\rho_{mix}:=I_S/d_S$ is a completely mixed state on $\Hi_S$.
\hfill $\blacksquare$

Note that $\Delta E_{SA}$ is evaluated with respect to the state before the SWAP between $S$ and
$a_1$. During Step 3, the state $\rho_A := \Tr_{SE}P\left(\ket{\Psi_{SEA}}\right)$ does not
change, and stays as a projector on $\Hi_A$ (up to a proportionality constant) as a result of
$\mathcal{F}_\mathrm{LF}^A$. Thus, if
$\Delta E_{SA} >0$, the Schmidt rank of $\ket{\Psi_{SEA}^{out}}$ is
strictly greater than that of $\ket{\Psi_{SEA}^{in}}$. This means that
$\textrm{rank}\rho_A^{out}-\textrm{rank}\rho_A^{in} \ge 1$; namely, applying the $SWAP_{Sa_1}$ in the
Step 1 increases the rank of the state on $A$ by at least 1.

Theorem \ref{sec:protocol_theorem_1} can now be proven with these lemmas and facts.\\
{\bf (Proof of Theorem \ref{sec:protocol_theorem_1})} \\
\quad The state on the joint system $SEA$ is always pure throughout the protocol as long as all
the local filtering operations succeed. Since the local filtering operation
$\mathcal{F}_\textrm{LF}^A$ preserves the Schmidt rank (increased due to time evolution) of
$\ket{\Psi_{SEA}}$ with respect to the partition $SE|A$, the execution of Step 1 increases the
Schmidt rank of $\ket{\Psi_{SEA}}$ by at least $1$ (as we have seen above). The Schmidt rank of
$\ket{\Psi_{SEA}}$ with respect to the same partition is obviously smaller than $d_E$. Thus $C$ is
no greater than $d_E$ and the statement of the theorem follows.
\hfill $\blacksquare$\\

The following lemma shows another, more intuitive, meaning of $\Delta E_{SA}$:
\begin{Lemma}\label{lemma:DeltaE=D+D}
The following equation holds for $\Delta E_{SA}$:
\begin{equation}
 \Delta E_{SA}= D(\rho_{SE}\| \rho_S \otimes \rho_E) + D(\rho_S \|\rho_{mix}),
\end{equation}
where $D(\rho \|\sigma )$ is the relative entropy defined as $D(\rho
 \|\sigma ) := \Tr \rho \log \rho -\Tr \rho \log
 \sigma$ \cite{NIELSEN,HAYASHI,WEHRL1978,OHYA,UMEGAKI1962,VEDRAL2002} and $\rho_{mix}=I_S/d_S$.
\end{Lemma}
({\bf Proof})
\begin{align*}
\quad & \Delta E_{SA}\\
=&  S(\rho _{E}) -  S(\rho_{SE}) + \log d_S \\
=& S(\rho _{E}) +S(\rho _S)- S(\rho _{SE}) - S(\rho_S)- \Tr\rho_S \log \rho_{mix} \\
=& I _\rho (S \| E) + D(\rho _S|\rho_{mix})\\
=& D(\rho_{SE}\| \rho_S \otimes \rho_E) + D(\rho_S \|\rho_{mix}),
\end{align*}
where $I_\rho (S \|E)):= S(\rho_S)+S(\rho_E)-S(\rho_{SE})$ is the mutual information between $S$
and $E$, and we use the formula $I_\rho(S\|E) = D(\rho _{SE}\|\rho_S \otimes \rho_E)$
\cite{NIELSEN} in the third equation. \hfill $\blacksquare$

This lemma guarantees that when $\Delta E_{SA}$ is small, $\rho_{SE}$ is close to
$\rho_S\otimes\rho_E$ and $\rho_S$ is almost completely mixed. We can also show the following:
\begin{Lemma}\label{lemma_Psi_SEA=Phi_SA_1_otimes_Phi_EA_2}
When $\rho_{SE}= \rho_S\otimes \rho_E$, there exists a decomposition of system $A$ into two
distinct subsystems $A_1$ and $A_2$; that is, $\Hi_A = \Hi_{A_1} \otimes \Hi_{A_2}$. Accordingly,
a (pure) state $\ket{\Psi_{SEA}}$ can be written as a product of pure states $\ket{\Phi_{SA_1}}
\in \Hi_S\otimes\Hi_{A_1}$ and $\ket{\Phi_{EA_2}}\in \Hi_E\otimes\Hi_{A_2}$ such that
\begin{equation}
\ket{ \Psi_{SEA}}= \ket{\Phi_{SA_1}} \otimes \ket{\Phi_{EA_2}},
\end{equation}
where $\ket{\Phi_{SA_1}}$ and $\ket{\Phi_{EA_2}}$ satisfies $\Tr _{A_1}P\left( \ket{\Phi_{SA_1}}
\right) =\rho_S$ and $\Tr _{A_2}P\left(\ket{\Phi_{EA_2}} \right) =\rho_E$, respectively. In
particular, when $\rho_S= \rho_{mix}$, $\ket{\Phi_{SA_1}}$ can be chosen as a standard maximally
entangled state, i.e., $\ket{\Upsilon_{SA_1}}=d_S^{-1/2}\sum_{i=1}^{d_S}\ket{i_S}\ket{i_{A_1}}$.
\end{Lemma}
({\bf Proof})\\
Consider any decomposition of $A$ into subspaces $A^\prime_1$ and $A^\prime_2$ such that $\Hi_A=
\Hi_{A^\prime_1} \otimes \Hi_{A^\prime_2}$ and $\dim \Hi_{A^\prime_1}=\dim \Hi_S$. Then there
exist the purifications $\ket{\Phi_{SA_1^\prime}}$ and $\ket{\Phi_{EA_2^\prime}}$ of $\rho_S$ and
$\rho_E$ in each subspace, i.e., $\Tr _{A_1^\prime} P(\ket{\Phi_{SA_1^\prime}})=\rho_S$ and $\Tr
_{A_2^\prime} P(\ket{\Phi_{EA_2^\prime}})=\rho_E$, respectively. As $\ket{\Psi_{SEA}}$ and
$\ket{\Phi_{SA^\prime_1}} \otimes \ket{\Phi_{EA^\prime_2}}$ have the same reduced density matrix
on $\Hi_S \otimes \Hi_E$, there exists a unitary operator $U_A$ on $\Hi_A$ such that
\cite{NIELSEN,HAYASHI}
\begin{equation}
\ket{\Psi_{SEA}} = U_A \ket{\Phi_{SA^\prime_1}}\otimes\ket{\Phi_{EA^\prime_2}}.
\end{equation}
Defining $\Hi_{A_1}\otimes \Hi_{A_2}:=U_A\Hi_{A^\prime_1}\otimes \Hi_{A^\prime_2}$ proves the
lemma. \hfill $\blacksquare$

With Theorem \ref{sec:protocol_theorem_1} and Lemmas \ref{lemma:DeltaE=D+D} and
\ref{lemma_Psi_SEA=Phi_SA_1_otimes_Phi_EA_2}, we finally arrive at the following theorem:
\begin{Theorem}\label{sec:protocol_theorem_2}
Let $K$ be the index of the counter $C$ when the protocol halts. Then, for all $t_\infty \ge t_K +
\Delta t$, $(d_E, \ket{\Psi_{SEA}(t_K)}, H_{SE})$ satisfies the ME condition for $[t_K,
t_\infty)$, where $\ket{\Psi_{SEA}(t_K)}$ is the state on $SEA$ after the $K$-th (successful)
application of the local filtering operation $\mathcal{F}_\mathrm{LF}^A$ in Step 2 at time
$t_K$.
\end{Theorem}

({\bf Proof })\\
By the construction of the protocol, $\Delta E_{SA}\left(t \right)=0$ for all $t \in \left[t_K,
t_K+ \Delta t\right]$. By Lemma \ref{lemma:DeltaE=D+D}, $\rho_{SE}(t)=\rho_S(t) \otimes \rho_E(t)$
and $\rho_S(t)$ is a maximally mixed state for all $t \in \left[t_K, t_k+ \Delta t\right]$. Then,
due to Lemma \ref{lemma_Psi_SEA=Phi_SA_1_otimes_Phi_EA_2}, for all $t \in [t_K, t_K + \Delta t]$,
there exist Hilbert spaces $\Hi_{A_1}(t)$ and $\Hi_{A_2}(t)$ such that $\Hi_A = \Hi_{A_1}(t)
\otimes \Hi_{A_2}(t)$ and $\ket{\Psi_{SEA}(t)}$ satisfies
\begin{equation}\label{sec:pme_subsec:pme_eq_def_pme4}
\Tr _E P\left( \ket{\Psi_{SEA}(t)} \right)= P\left( \ket{\Upsilon_{SA_1(t)}}
                       \right) \otimes \rho_{A_2(t)},
\end{equation}
where $\ket{\Upsilon_{SA_1(t)}}$ is a maximally entangled state on $\Hi_S\otimes\Hi_{A_1}(t)$.
Moreover, $\rho_{A_2(t)}(t)$ is a projector, because of the local filtering operation in Step 2.
Therefore, $\left( d_E, \ket{\Psi_{SEA}(t_K)}, H_{SE} \right)$ fulfills the ME condition for
$[t_K, t_K +\Delta t]$. Corollary \ref{sec:pme_subsec:pme_is_mes_corollary} then lets us finish
the proof of the theorem.
\hfill $\blacksquare$ \\

Therefore, steering the entire system of $S, E$, and $A$ towards one that satisfies the ME
condition can be achieved within a finite time. Once this state-steering has been done, the
identification of the equivalence class by probing only systems $S$ and $A$ can be performed, as
shown in Sec. \ref{sec:tomography}.

In the present study, we have assumed that the state tomography on $SA$ can be performed
perfectly. The feasibility of the presented protocol for state-steering depends on this
assumption: in order to complete the task, we need to check whether $\Delta E_{SA}(t)$ is exactly
$0$ for all $t \in \left[t_C, t_C + \Delta t \right]$.

\section{Numerical simulation of the protocol}\label{sec:numerical}
\begin{figure}
\includegraphics[scale=1]{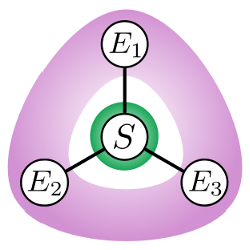}
\caption{(Color online) The spin network for the numerical simulation. The small circles denote
spins-1/2 and the edges connecting them represent interaction between spins. The central spin $S$
interacts with the three surrounding spins, $E=\{E_1, E_2, E_3\}$, according to the Hamiltonian
Eq. (\ref{eq_sec:numerical_Hamiltonian_SE}). }\label{fig:spinstar}
\end{figure}

In order to reinforce our idea with a concrete example, let us present the results of a numerical
simulation. Here, we consider a star-shaped network of spins-1/2 with the Heisenberg-type
inter-spin interaction. We focus on a four-spin network, and identify the central spin as $S$ and
three surrounding spins as $E$ (See Fig. \ref{fig:spinstar}). The whole system is represented by a
Hilbert space $\Hi_{SE}=\Hi_S \otimes \Hi_E$ with $\Hi _S = \mathbb{C}^2$ and $\Hi_E = \left(
\mathbb{C}^2 \right)^{\otimes 3}$, and the system Hamiltonian $H_{SE}$ is written as
\begin{equation}\label{eq_sec:numerical_Hamiltonian_SE}
 H_{SE} := \frac{1}{2}\sum_{(m,n)\in
\mathcal{E}} J_xX_m X_n+J_yY_mY_n+J_zZ_mZ_n,
\end{equation}
where $\mathcal{E}$ is the set of edges connecting spins $m$ and $n$ in the network.

In the simulation, we set the coupling constants to be $J_x=J_y=J_y=1$ for all edges, and we use
the eigenbasis, $\{\ket{0},\ket{1}\}$, of the standard Pauli $Z$ operator as the basis for each
spin, i.e., $Z\ket{i}=(-1)^i\ket{i} \; (i=\{0,1\})$. We will consider three types of initial
states, $\ket{\Psi_{1}}$, $\ket{\Psi_{2}}$, and $\ket{\Psi_{3}}$, which are defined as
\begin{align}
\ket{\Psi_{1}}&= \ket{0}_S \ket{e_1}_E, \label{sec:numerical_eq_def_Psi1}\\
\ket{\Psi_{2}} &= \frac{1}{\sqrt{2}} \ket{0}_S \left(\ket{e_1}_E
 +\ket{e_2}_E \right), \label{sec:numerical_eq_def_Psi2} \\
\ket{\Psi_{3}} &= \frac{1}{2}\ket{0}_S \left(\ket{e_1}_E
 +\ket{e_2}_E +2\ket{e_3}\right). \label{sec:numerical_eq_def_Psi3}
\end{align}
The three state vectors on $\Hi_E$, $\ket{e_0}$, $\ket{e_1}$, and $\ket{e_2}$, are given by
\begin{align}
 \ket{e_1} & :=  -\frac{2}{\sqrt{6}}\ket{100}+\frac{1}{\sqrt{6}}\left(\ket{010}+\ket{001}
 \right),\\
 \ket{e_2} & := \frac{1}{\sqrt{2}}\left(\ket{110}-\ket{101}\right), \\
 \ket{e_3} & := \ket{000},
\end{align}
and these $\ket{e_i}$ are orthogonal to each other. Note that the reduced density matrices on $S$
of these states are all $\ket{0}\bra{0}$, thus the differences between $\ket{e_i}$ cannot be seen
from an observer. The reason for these choices of the initial states will be clear later.

\begin{figure}
\includegraphics[scale=0.46]{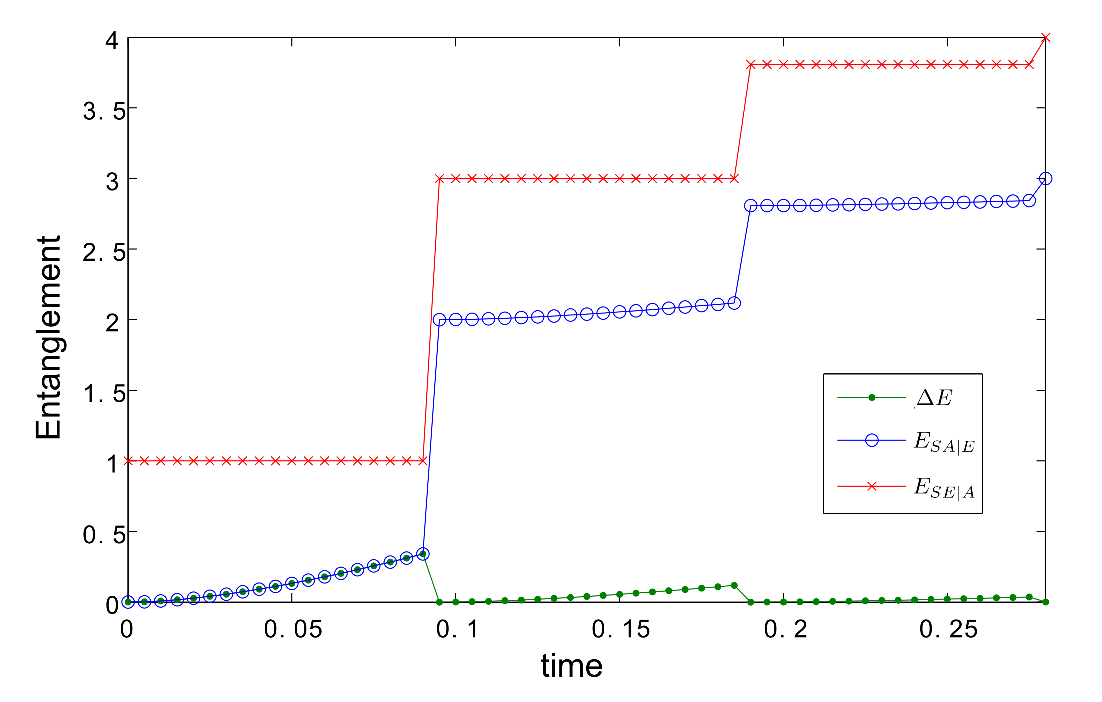}
\caption{(Color online) The changes in $\Delta E$, $E_{SE|A}$, and $E_{SA|E}$ by the
state-steering protocol as
 functions of time $t$. The initial state is set to be $\ket{\Psi_3}$ defined by Eq. (\ref{sec:numerical_eq_def_Psi3}).
 The green line with dots ($\cdot$), the blue line with circles ($\circ$), and the red line with crosses ($\times$),
 represent $\Delta E$, $E_{SA|E}$, and $E_{SE|A}$, respectively. $\Delta E$ is given by $\Delta E := E_{SA|E} - E_{SE|A}+1$
 according to Eq. (\ref{Delta_E_SA}).}\label{fig:sim_entanglement_changes}
\end{figure}

Figure \ref{fig:sim_entanglement_changes} shows the simulated time evolution of $\Delta E$ defined
by Eq.~(\ref{Delta_E_SA}), $E_{SE|A}$, and $E_{SA|E}$, where $E_{X|Y}$ denotes the amount of
entanglement between subsystems $X$ and $Y$. The initial state for this plot was chosen to be
$\ket{\Psi_3}$. In this particular run of the simulation, the local filtering operations in Step 2
of the state steering protocol (cf. Sec. \ref{sec:state_steering}) were performed at times $t=0$,
$0.095$, $0.19$, and $0.28$. As we have mentioned in Section \ref{sec:state_steering}, $E_{SE|A}$
grows monotonically: it stays constant except for the instances when the local filtering is
successfully executed. We also observe that $E_{SA|E}$ seems to increase monotonically as well,
although it does so only because we set the parameter $\Delta t_C$ sufficiently small.

We see in Fig.~\ref{fig:sim_entanglement_changes} that the final state $\ket{\Phi_3}$ of the
state-steering protocol satisfies $E_{SE|A}\left( \ket{\Phi_3} \right)=4$, which means its Schmidt
rank is $16$. (We let $\ket{\Phi_i}$ denote the final state of the protocol after starting with
$\ket{\Psi_i}$.) We have also simulated the protocol with the initial states $\ket{\Psi_1}$ and
$\ket{\Psi_2}$. The results indicate that the Schmidt ranks of the corresponding final states
$\ket{\Phi_1}$ and $\ket{\Phi_2}$ are $4$ and $8$, respectively (plots not shown). Therefore,
$\ket{\Phi_3}$ is a maximally entangled state in the sense that its support fully covers the
entire Hilbert space $\mathcal{H}_S\otimes \mathcal{H}_E$, while $\ket{\Phi_1}$ and $\ket{\Phi_2}$
are not, although all of them satisfy the ME condition.

As we have described in Sections \ref{sec:pme} and \ref{sec:tomography}, the Schmidt rank of the
final state of the state-steering protocol is identified as $d_S\tilde{d}_E$ with $\tilde{d}_E$
being the estimated dimension of $E$. Suppose $\tilde{d}_E^{(i)}$ is the estimated dimension after
going through the protocol with initial state $\ket{\Psi_i} \; (i=\{1,2,3\})$. Then, we have
$\tilde{d}_E^{(1)}=2$, $\tilde{d}_E^{(2)}=4$, and $\tilde{d}_E^{(3)}=8$. Although the three
initial states lead to the same reduced density matrix $\ket{0}\bra{0}$ on $S$, these examples
clearly show that an estimated dimension of $E$ depends on the state $\rho_E$ of the system $E$.
The details of the classification of states will be discussed in the near future \cite{OMK12}.

We have also carried out a numerical simulation of the tomography part of the protocol in Sec.
\ref{sec:tomography}. Time $t$ is reset to zero at the instance when the state steering protocol
is terminated. Then, for each output state $\ket{\Phi_i} \;(i=\{1,2,3\})$ of the state-steering
protocol we computed its time evolution under $H_{SE}$ and generated a sequence of reduced density
matrices $\rho_{SA}(t)$ from $t=0$ to $t=n \cdot dt$ with interval $dt$. We chose $n=1,500,000$
and $dt=0.1 \cdot \pi/\|H_{SE}\|_{op} \approx 0.063$. In Section \ref{sec:tomography}, we obtained
$\left\{ \theta _\alpha \right\}_{\alpha=1}^L$ and $\left\{ \rho _\alpha \right\}_{\alpha=1}^L$
defined in Eq. (\ref{sec:pme_subsec:tomography_eq_rhoSA_t}) from Eq.
(\ref{sec:pme_subsection_tomography_eq_dn_dt_n_rhoSA}). However, since Eq.
(\ref{sec:pme_subsection_tomography_eq_dn_dt_n_rhoSA}) involves higher order derivatives, this
method is not preferable for numerical calculation. Therefore, here, we first performed Fourier
transform of the data of matrix elements to get $\left\{ \theta _\alpha \right\}_{\alpha=1}^L$,
which appear as the positions of peaks. Then, we derived $\left\{ \rho _\alpha
\right\}_{\alpha=1}^L$ by solving Eq. (\ref{sec:pme_subsec:tomography_eq_rhoSA_t}), which is a
system of linear equations about $\left\{ \rho _\alpha \right\}_{\alpha=1}^L$ for given $\left\{
\theta _\alpha \right\}_{\alpha=1}^L$ and matrix elements of $\rho_{AS}(t)$.

Finally, by solving the system of linear equations (\ref{system_of_equations_for_hj}), using the
values $\left\{ \theta _\alpha \right\}_{\alpha=1}^L$ and $\left\{ \rho _\alpha
\right\}_{\alpha=1}^L$, we estimated a Hamiltonian $\tilde{H}_{SE}^{(i)}$ corresponding to each
initial state $\ket{\Phi_i}$. Note that, as a consequence of the differences between
$\tilde{d}_E^{(i)}$, the estimated Hamiltonian $\tilde{H}_{SE}^{(i)}$ is a hermitian operator on
$\tilde{d}_E^{(i)}$-dimensional space.

As we have discussed in Sec. \ref{sec:pme}, there are multiple (in fact infinite) possibilities
for the triple $(d_E, \ket{\Psi_{SEA}}, H_{SE})$ that lead to indistinguishable dynamics on $SA$,
no matter what operations we perform on $SA$. Therefore, what we can estimate by the method above
is one possible Hamiltonian among those in the same equivalence class, and it is likely that the
estimated Hamiltonian looks very different from the \textit{true} Hamiltonian, Eq.
(\ref{eq_sec:numerical_Hamiltonian_SE}). Let us now verify that despite apparent differences
between them the observable time evolution of the state $\rho_{SA}$ is indeed the same
irrespective of the choice of Hamiltonian in the equivalence class.

Since we have identified three types of dynamics, depending on the initial state $\ket{\Psi_i}$,
let us call the corresponding estimated Hamiltonian $\tilde{H}_{SE}^{(i)}$ with $i=\{1,2,3\}$. The
matrix elements of the first three rows and columns of the true Hamiltonian $H_{SE}$ are
\begin{equation}\label{true_hamiltonian}
 H_{SE} = \frac{1}{2}\left(
\begin{array}{cccc}
3 & 0 & 0 & \cdots \\
0 & 1 & 0 & \cdots \\
0 & 0 & 1 & \cdots \\
\vdots & \vdots & \vdots & \ddots
\end{array}
\right),
\end{equation}
where the basis is taken as $\{\ket{0000},\ket{0001},\ket{0010},...\}$. The same part of the
estimated Hamiltonians are
\begin{equation}\label{est_hamiltonian1}
\tilde{H}_{SE}^{(1)} = \left(
\begin{array}{cccc}
0 & 0 & 0 & \cdots \\
0 & 0.167 & 0.167 & \cdots \\
0 & 0.167 & 0.167 & \cdots \\
\vdots & \vdots & \vdots & \ddots
\end{array}
\right),
\end{equation}

\begin{equation}\label{est_hamiltonian2}
\tilde{H}_{SE}^{(2)} = \left(
\begin{array}{cccc}
0 & 0 & 0 & \cdots \\
0 & 1.167 & -1.441 + 0.182i & \cdots \\
0 & 1.441- 0.182i & 1.274 & \cdots \\
\vdots & \vdots & \vdots & \ddots
\end{array}
\right),
\end{equation}

\begin{equation}\label{est_hamiltonian3}
\tilde{H}_{SE}^{(3)} = \left(
\begin{array}{cccc}
3.952 & 0 & 0 & \cdots \\
0 & 1.372 & 0.093+0.072i & \cdots \\
0 & 0.093+0.072i & 0.536 & \cdots \\
\vdots & \vdots & \vdots & \ddots
\end{array}
\right).
\end{equation}

What interests us is the difference between the effect of $H_{SE}$ and that of $\tilde{H}_{SE}$.
For this comparison, we computed the two states on $SEA$ at time $t$, one is driven by $H_{SE}$
and the other is by $\tilde{H}_{SE}^{(i)}$ with the corresponding initial state $\ket{\Phi_i}$
\footnote{The state $\ket{\Phi_i}$ is the final state of the state-steering protocol, and is now
used as the initial state to see the time evolution caused by $\tilde{H}_{SE}^{(i)}$ as well as
$H_{SE}$.}. Namely, $\ket{\Phi_{SEA}^{(i)}(t)}=\exp(-iH_{SE}t)\ket{\Phi_i}$ and
$\ket{\tilde{\Phi}_{SEA}^{(i)}(t)}=\exp\left(-i\tilde{H}_{SE}^{(i)}t\right)\ket{\Phi_i}$, from
which reduced density matrices $\rho_{SA}^{(i)}(t)$ and $\tilde{\rho}_{SA}^{(i)}(t)$ are obtained.

\begin{figure}
\includegraphics[scale=0.80]{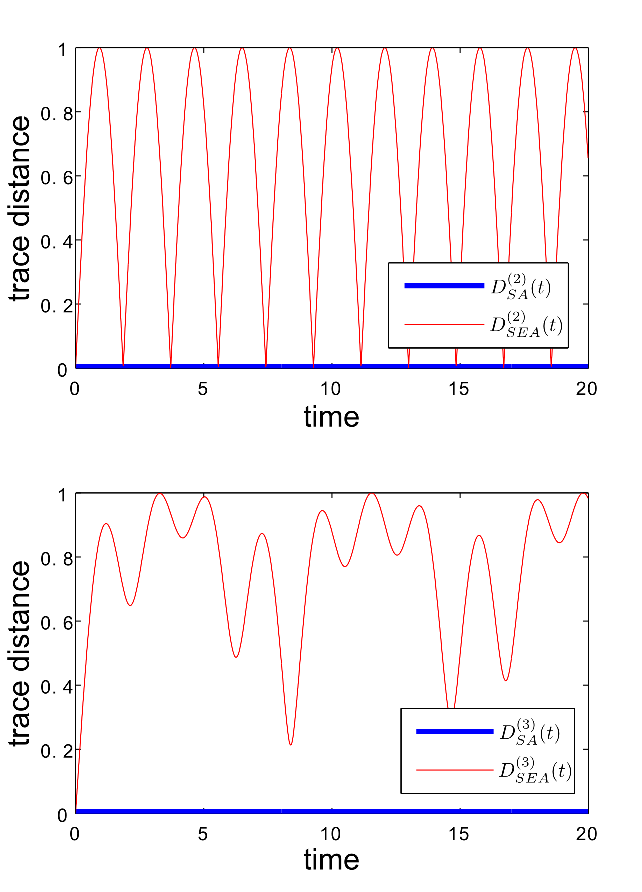}
\caption{(Color online) The comparison of the time evolution of the state $\ket{\Phi_{SA}^{(i)}}$
and $\ket{\Phi_{SEA}}$ under the two Hamiltonians, $H_{SE}$ and $\tilde{H}_{SE}^{(i)}$. (a) is for
the case of $i=2$ and (b) is for $i=3$. The blue thick line and the red thin line represent the
trace distances $D_{SA}^{(i)}(t)$ and $D_{SEA}^{(i)}(t)$, respectively.}
\label{fig:sim_trace_distance}
\end{figure}

Figure \ref{fig:sim_trace_distance} shows how the states $\rho_{SA}$ and $\rho_{SEA}$ are affected
by those Hamiltonians. We measure the difference between states with trace distance, and plot it
in the figure. For two states on $SA$, $\rho_{SA}^{(i)}(t)$ and $\tilde{\rho}_{SA}^{(i)}(t)$, the
trace distance is defined as
\begin{equation}\label{def_trace_distance}
D_{SA}^{(i)}(t) := \frac{1}{2} \| \rho_{SA}^{(i)}(t)-\tilde{\rho}_{SA}^{(i)}(t)\|_\mathrm{tr}
\end{equation}
where $\| A-B\|_\mathrm{tr}=\Tr\sqrt{(A-B)^\dagger (A-B)}$. The distance $D_{SEA}^{(i)}(t)$
between $\ket{\Phi_{SEA}^{(i)}(t)}$ and $\ket{\tilde{\Phi}_{SEA}^{(i)}(t)}$ is defined similarly.

In Fig.~\ref{fig:sim_trace_distance}(a), $D_{SA}^{(i)}(t)$ and $D_{SEA}^{(i)}(t)$ are plotted for
the case of $i=2$, and Fig. \ref{fig:sim_trace_distance}(b) is for $i=3$. The plot for the $i=1$
case is not shown here because both $D_{SA}^{(1)}(t)$ and $D_{SEA}^{(1)}(t)$ stay zero for all
$t\ge 0$.

In both Figs. \ref{fig:sim_trace_distance}(a) and (b), we can see that the two different-looking
Hamiltonians, $H_{SE}$ and $\tilde{H}_{SE}^{(i)}$, give rise to the identical dynamics on $SA$,
i.e., $D_{SA}^{(i)}(t)=0$, while the entire state on $SEA$ evolves quite differently under these
Hamiltonians. This observation convincingly confirms the equivalence we have analysed in this
work, that is $(d_E, \ket{\Phi_i}, H_{SE})\equiv (\tilde{d}_E^{(i)}, \ket{\Phi_i},
\tilde{H}_{SE}^{(i)})$.

The three states defined in Eqs.
(\ref{sec:numerical_eq_def_Psi1})-(\ref{sec:numerical_eq_def_Psi3}), $\ket{\Psi_i} \;
(i=\{1,2,3\}$ showed distinct behaviours under the Hamiltonian $\tilde{H}_{SE}^{(i)}$ that was
estimated to describe the observable dynamics on $SA$. The classification of the states, or more
precisely subspaces, in $\Hi_S\otimes\Hi_E$ is an interesting subject in its own right, hence we
will present it separately in the near future \cite{KMO13}.

To summarize, we have performed numerical simulations of our protocol, focusing on the Heisenberg
model with a star-shaped graph. We considered three different initial states $\ket{\Psi_i} \;
(i=1, 2, 3)$ for the state-steering protocol as in Eqs.(\ref{sec:numerical_eq_def_Psi1}) -
(\ref{sec:numerical_eq_def_Psi3}). Then, the estimated dimension $\tilde{d}_E^{(i)}$ of $E$ turned
out to be different, depending on $\ket{\Psi_i}$. Despite such a nontrivial difference, we have
confirmed that all estimated Hamiltonians $\tilde{H}_{SE}^{(i)} \; (i=1, 2, 3)$ generate a time
evolution on system $SA$, which is identical to that we expect from the `true' Hamiltonian
$H_{SE}$. These results suggest that our method can extract minimal, but sufficient, information
on the system Hamiltonian to account for the dynamics on $SA$. Finally, we note that in order to
explain what causes the differences in $D_{SA}^{(i)}(t)$, we need to analyze the dynamical Lie
algebra for the whole system more deeply, and it will be presented in a follow-up paper
\cite{KMO13}.

\section{Conclusion}\label{sec:conclusion}

We have shown the possibility of probing a large surrounding quantum system (environment) through
a small principal system, provided the environment is effectively finite-dimensional and the
entire system can be initialised to be a fixed (unknown) state. By probing, we mean fully
identifying the Hamiltonian for the purpose of utilising it as a useful resource for quantum
control, e.g., quantum computation.

In analyzing our idea, we have found that there are equivalence classes, in terms of the observable dynamics, induced by the limited access. We have also found that the fulfillment of a condition, i.e.,  the
\textit{ME condition}, which is verifiable without a direct access to the environment, is sufficient for the reconstruction of a representative of the equivalence class (Sec. \ref{sec:pme}). This reconstruction can be achieved
by performing state tomography of the joint system $S$ and $A$ without active operations on them (Sec.~\ref{sec:tomography}).  In order to make the state satisfy the ME condition, we have constructed a protocol to steer the entire system (Sec. \ref{sec:state_steering}). Determining the equivalence class
provides us with the full information on the Hamiltonian $H_{SE}$ to indirectly control the
environment. As a concrete corroboration of our theoretical analyses, we have carried out a
numerical simulation in Sec. \ref{sec:numerical}.

Although we have focused on the theoretical
aspect of our tomographic scheme for a `not directly probable system', which is quite remarkable
in its own right, any quantum operation, including state tomography, is always fraught with the
effect of unpredictable noise in reality. Therefore, from a pragmatic point of view, we need to
modify the protocol and evaluate errors that may occur in the equivalent class identification.
Since the analysis of errors in the protocol is beyond the scope of this paper, we leave it as a
future project \cite{OMK12}.

Considering the extreme difficulty of manipulating a huge number of individual quanta, indirect
control seems a rational approach. While the experimental implementation of the protocol might be unrealistic today, our method opens up a path to the novel exploitation of high dimensional
quantum systems with minimal artificial controls.

\section*{Acknowledgements}
We thank Kiyoshi Tamaki, Koji Azuma and Fernando Brand\~{a}o for useful discussions. KM and TT are
supported in part by Quantum Cybernetics (Grant No. 2112004), CREST-JST, and FIRST-JSPS (Quantum
Information Process).

\bibliographystyle{apsrev}

\end{document}